\RequirePackage{fix-cm}
\documentclass[twocolumn,epjc3]{svjour3}  
\usepackage{graphicx} 
\usepackage{dcolumn}
\usepackage{bm}
\usepackage{subfigure}
\usepackage{float}
\usepackage{multirow}
\usepackage{booktabs}
\usepackage{amsmath}
\usepackage{cases}
\usepackage{latexsym,amssymb}
\usepackage[mathscr]{eucal}
\usepackage[switch]{lineno}
\usepackage{array}
\usepackage{comment}
\usepackage[bookmarks=true,
   colorlinks=true,
   linkcolor=blue,
   urlcolor=blue,
   citecolor=blue,
   bookmarks=true,
   hyperindex=true
]{hyperref}
\smartqed  

\journalname{Eur. Phys. J. C}
\begin{document}

\title{Global curvature control in a regular black hole with Minkowski core and inner-extremal horizon}


\author{Zhong-Wen Feng\thanksref{e1,addr1}
        \and
        Hong-Lin Liu\thanksref{e2,addr1}
        \and
        Yi Ling\thanksref{e3,addr2,addr3}
        \and
        Qing-Quan Jiang\thanksref{e4,addr1}
}

\thankstext{e1}{e-mail: zwfengphy@cwnu.edu.cn}
\thankstext{e2}{e-mail: hlliuphy@163.com}
\thankstext{e3}{e-mail: lingy@ihep.ac.cn}
\thankstext{e4}{e-mail: qqjiangphys@yeah.net}


\institute{School of Physics and Astronomy,
China West Normal University, Nanchong 637009, China \label{addr1}
           \and
           Institute of High Energy Physics, Chinese Academy of Sciences, Beijing 100049, China \label{addr2}
           \and
           School of Physics, University of Chinese Academy of Sciences, Beijing 100049, China \label{addr3}
}

\date{Received: date / Accepted: date}

\maketitle
\begin{abstract}
We construct a static, spherically symmetric regular black hole with a Minkowski core, a non-extremal outer horizon, and a degenerate inner horizon with vanishing surface gravity. The spacetime exhibits two main features. First, in the large-mass regime with $r_+=2M$, the maximal Kretschmann scalar becomes nearly independent of the ADM mass and is mainly controlled by the inner-horizon radius $r_-$, providing a simple sufficient condition for the curvature to remain sub-Planckian everywhere. Second, the effective cubic zero of the metric function replaces the standard exponential near-horizon amplification with power-law behavior. A massless scalar test field shows that the geometric blueshift factor grows as $v^{3/2}$. Then, we use the double-null shell and Ori models to examine the corresponding effective mass dynamics. the nonlinear shell-crossing contribution decays at late times, while the Ori evolution with a Price-law ingoing flux keeps the internal Misner-Sharp mass bounded. In both descriptions, the late-time mass approaches $r_-/2$. The same inner-horizon scale therefore controls the global curvature bound and the finite mass scale reached by the shell dynamics.

\end{abstract}

\section{Introduction}
\label{intro}
Classical general relativity predicts that black holes contain the singularities of spacetime~\cite{Penrose:1964wq,Hawking:1976ra}. This is usually regarded as an indication that the classical description breaks down in the strong-field regime. It is therefore natural to ask whether one can construct effective black hole geometries in which the singular interior is removed. Regular black holes (RBHs) provide one possible realization of this idea and have been studied for many years as effective models of singularity resolution. Such geometries may arise from effective matter sources, nonlinear electrodynamics, or quantum-gravity-inspired corrections~\cite{Bardeen1968,Dymnikova:1992ux,AyonBeato:2000zs,Hayward2006,SpallucciSmailagic2017,Simpson:2018tsi,Ashtekar:2018cay,Burzilla:2020utr,Torres2023,LanMiao2023,Feng:2023pfq,Bambi:2023try,Bueno:2024dgm,Frolov:2024hhe,Ling:2025ncw,Liu:2026zzd}. In these models, the Schwarzschild behavior is usually recovered at large radii, while the deep interior is modified to avoid the central singularity.

Nevertheless, removing the central singularity is only part of the problem~\cite{Carballo-Rubio:2025fnc}. Most RBHs also contain an inner horizon, which is known to be sensitive to perturbations. In the standard scenario, ingoing and outgoing radiations are strongly blueshifted near the inner (Cauchy) horizon, leading to rapid growth of the effective internal mass~\cite{Carballo-Rubio:2019nel,Carballo-Rubio:2019fnb}. This phenomenon is known as mass inflation~\cite{PoissonIsrael1989,PoissonIsrael1990,Ori1991,Brady:1995ni,Bertipagani:2020awe,Brown:2011tv,Hale:2025ezt}. Its physical mechanism and nonlinear development have been discussed extensively, see e.g., Refs.~\cite{Hamilton:2008zz,Hwang:2010im} and the references therein. Thus, the construction of a viable RBH requires not only the removal of the central singularity but also a controlled behavior of the inner horizon.

This issue has become increasingly important in recent studies of RBHs. In Ref.~\cite{CarballoRubio2018}, it is emphasized that the absence of a central singularity does not guarantee the absence of internal instabilities. Subsequent analyses show that non-perturbative backreaction effects near the inner horizon may significantly modify the geometry and challenge the validity of the semiclassical description~\cite{CarballoRubio2021,DiFilippo2022}. On the other hand, regular black holes with stable cores have been discussed in a dynamical framework, suggesting that the inner region may remain regular under suitable conditions~\cite{Bonanno2021}. The interplay between mass inflation and semiclassical backreaction near inner horizons is also investigated in Refs.~\cite{Barcelo2021,Barcelo2022}. More recently, fully nonlinear simulations of scalar-field collapse in the Hayward geometry have shown that the fate of the inner horizon depends sensitively on the strength of the perturbation, with weak scalar fields leaving a finite-radius inner horizon and sufficiently strong perturbations driving the inner horizon to zero and leading to the formation of a spacelike singularity~\cite{Shao:2025apr，Shao:2025fki}. These studies indicate that the stability of the inner horizon is one of the central consistency tests for RBH models.

An important step in this direction was made in Ref.~\cite{CarballoRubio2022}, where explicit RBHs with a degenerate inner horizon and a non-extremal outer horizon were constructed. Since $\kappa_-=0$, the standard exponential amplification responsible for mass inflation can be softened, as demonstrated there using both the double-null shell and Ori models. The inner-extremal construction was subsequently extended to rotating RBHs and realized in higher-curvature pure gravity and modified gravity~\cite{Franzin2022,DiFilippo2024mwm,Frolov:2026rcm,Eichhorn:2025pgy}. Related studies have explored both situations in which inner-horizon growth may be limited~\cite{Bonanno:2022rvo,bonanno2025cauchyhorizoninstabilityregular} and residual semiclassical or dynamical effects that can persist even when the standard exponential amplification is suppressed~\cite{McMaken2023,Bonanno:2023rzk,Khodadi:2024efq,Carballo-Rubio:2024dca}. These results show that the near-inner-horizon geometry plays a central role in determining the physical viability of RBHs. 

These developments leave two related model-dependent issues unresolved. Central regularity requires curvature invariants to remain finite, but it does not ensure that their maxima are controlled by a fixed microscopic scale. Since Planck-scale curvature marks a regime in which higher-curvature and quantum corrections are no longer guaranteed to remain suppressed, this distinction motivates the limiting-curvature condition, which imposes uniform curvature bounds set by a fundamental length~\cite{Ashtekar:2018cay,Burgess:2003jk,Frolov:2016pav}. The near-horizon dynamics of a specific inner-extremal geometry must also be examined, since the vanishing of $\kappa_-$ softens the standard exponential blueshift but does not directly determine the late-time evolution of the internal mass~\cite{CarballoRubio2022}.

To address these issues, we construct a static, spherically symmetric regular black hole with an asymptotically Schwarzschild exterior, a Minkowski core, a non-extremal outer horizon, and a degenerate inner horizon with vanishing surface gravity. The aim is to determine whether a single explicit geometry can combine the softened near-inner-horizon amplification with a globally controlled curvature scale. The nonpolynomial auxiliary function $g\left(r\right)$ provides such a realization while preserving an effective cubic zero at the inner horizon. In the large-mass regime with $r_+=2M$, the maximal Kretschmann scalar becomes nearly independent of the ADM mass and is mainly controlled by the independent inner-horizon scale $r_-$, yielding a simple sufficient condition for the curvature to remain sub-Planckian everywhere. Treating $r_-$ as a fixed scale also has a model-specific consequence for the mass response. When neighboring geometries with different ADM masses are compared at fixed $r_-$, the quadratic response associated with a mass-dependent inner-horizon position is absent, and the leading response becomes $\partial m/\partial M=\mathcal{O}\left[\left(r-r_-\right)^3\right]$. Together with the near-horizon scaling $r-r_-\sim v^{-1/2}$, this cubic response makes the nonlinear DTR contribution decay as $v^{-3/2}$, while the Ori evolution with a Price-law ingoing flux keeps the internal Misner-Sharp mass bounded. The connection between the global curvature control and the cubic near-horizon mass response constitutes the model-specific result of the present construction.

The outline of this paper is as follows. In Sec.~\ref{sec2}, we construct the regular black hole spacetime and analyze its horizon structure, global curvature scale, and effective energy conditions. In Sec.~\ref{sec3}, we first use a massless scalar test field to derive the local near-inner-horizon scaling. We then examine the fixed-$r_-$ mass response and the nonlinear shell-crossing term in the double-null shell model, followed by the internal mass evolution in the Ori model with a Price-law ingoing flux. We also compare the resulting behavior with representative regular black hole constructions. The conclusions and discussion are presented in Sec.~\ref{sec4}.  Throughout this paper, we use units in which $G=\ell_{\rm p}^{2}=1$.

\section{Construction and spacetime properties of the regular black hole}
\label{sec2}
\subsection{Construction of the metric function}
\label{sec2-1}
We consider a static, spherically symmetric spacetime of the following form
\begin{align}
\label{eq1}
{\rm{d}}{s^2} =  - f\left(r \right){\rm{d}}{t^2} + f{\left( r \right)^{ - 1}}{\rm{d}}{r^2} + {r^2}{\rm{d}}{\Omega ^2},
\end{align}
where $f\left(r \right)$ is the metric function to be determined.

In the present work, our aim is to construct a new RBH spacetime that satisfies the following requirements. First, the spacetime should be asymptotically Schwarzschild-like at large distances. Second, the central region should remain regular as $r\to0$. Third, the spacetime should possess a simple outer horizon at $r=r_+$ and a degenerate inner horizon at $r=r_-$, namely $\kappa_+\neq0$ and $\kappa_-=0$. Fourth, the curvature scale of the spacetime should admit a parameter regime in which the Kretschmann scalar remains below the Planck scale everywhere. To meet these requirements, the metric function can be written as
\begin{align}
\label{eq2}
f\left( r \right) = 1 - \frac{{2 m\left( r \right)}}{r} =\frac{{{N}\left( r \right)}}{{{D}\left( r \right)}},
\end{align}
where ${m\left( r \right)}$ is the Misner-Sharp mass, the numerator and the denominator are chosen according to the desired horizon structure, asymptotic behavior, and regularity conditions.

A key point of the construction in Ref.~\cite{CarballoRubio2022} is that the standard exponential amplification associated with mass inflation can be avoided if the inner horizon is sufficiently degenerate so that its surface gravity vanishes. In the present work, it is worth noting that the root of $f\left(r\right)$ at $r=r_-$, namely $f\left(r_-\right)=0$, should be of odd order. Since the outer horizon $r=r_+$ is required to be a simple root, the metric function changes its sign when crossing $r=r_+$ and becomes negative in the trapped region $r_-<r<r_+$. On the other hand, regularity at the center requires the metric function to become positive again as $r\to0$. Thus, in order for the metric function to recover the correct sign inside the core, the zero at $r=r_-$ must be of odd order, and the lowest admissible choice is an effective triple root. A double zero at $r=r_-$ is therefore not sufficient, even though it already gives $\kappa_-=0$. However, the essential requirement is not that the numerator must be a polynomial in $\left(r-r_-\right)$, but rather that its local behavior near the inner horizon should satisfy $N\left( r \right) \sim {\left( {r - {r_ - }} \right)^3}\left( {r - {r_ + }} \right)$. In other words, what really matters is the effective order of the zero at $r=r_-$, rather than the explicit polynomial form adopted in Ref.~\cite{CarballoRubio2022}. This observation allows us to generalize the construction and write the numerator as
\begin{align}
\label{eq3}
N\left( r \right) = g\left( r \right)\left( {r - {r_ - }} \right)\left( {r - {r_ + }} \right),
\end{align}
provided that the auxiliary function satisfies $g\left( r \right) \sim {\left( {r - {r_ - }} \right)^2}$ as $r\to r_-$. Under this form, the numerator still has the same effective behavior $N\left( r \right) \sim {\left( {r - {r_ - }} \right)^3}\left( {r - {r_ + }} \right)$  near the inner horizon and therefore yields the same vanishing inner horizon surface gravity. In addition, this choice modifies the spacetime in the region $r<r_-$ and helps keep the overall curvature scale under better control. Inspired by the exponential-type modifications used in Refs.~\cite{Xiang:2013sza,LingLingYi:2021rfn,Ling:2021olm}, we choose
\begin{align}
\label{eq4}
g\left( r \right) = \exp {\left( {1 - \frac{{{r_ - }}}{r}} \right)^2} - 1.
\end{align}
It is easy to see that $g\left( r \right)$ has a double zero at $r=r_-$. As a result, the numerator in Eq.~(\ref{eq3}) has an effective triple zero at the inner horizon. At the same time, the exponential factor modifies the spacetime inside the inner horizon and helps keep the curvature scale under control.

Next, we construct the denominator, which should preserve the root structure determined by the numerator while ensuring the correct asymptotic and central behavior. More specifically, it should satisfy three basic requirements.  First, it should be finite and nonzero at both $r=r_+$ and $r=r_-$, with the multiplicity of the two horizons therefore still determined by the numerator. Second, in the large-$r$ region, the metric function should reproduce the Schwarzschild asymptotic form. Third, the denominator should have the same dominant behavior as the numerator near the center, so that the metric function remains regular. Since $g\left(r\right)\to e-1$ as $r\to\infty$, a simple choice satisfying these requirements is
\begin{align}
\label{eq5}
D\left(r\right)=N\left(r\right)+2\left(e-1\right) M r.
\end{align}

Indeed, at the two horizons $D\left(r_\pm \right)=2 \left(e-1 \right)M r_\pm\neq0$, which means that the denominator does not alter the multiplicity of the roots. For $0<r_-<r_+=2M$, one has $D\left(r\right)>0$ for all $r>0$, so the denominator introduces no additional finite-radius poles. The second term in Eq.~(\ref{eq5}) is responsible for recovering the Schwarzschild behavior at large distances. Moreover, as $r\to0$, the function $g\left(r\right)$ grows rapidly, and the numerator and denominator are dominated by the same leading term. Therefore, the metric function takes the form
\begin{align}
\label{eq6}
f\left( r \right)& = \frac{{g\left( r \right)\left( {r - {r_ - }} \right)\left( {r - {r_ + }} \right)}}{{g\left( r \right)\left( {r - {r_ - }} \right)\left( {r - {r_ + }} \right) + 2\left( {e - 1} \right)Mr}},
\nonumber\\
g\left( r \right) &= \exp {\left( {1 - \frac{{{r_ - }}}{r}} \right)^2} - 1.
\end{align}
Now, the metric function is  fully specified. In the next subsection, we analyze the horizon structure, asymptotic behavior, and curvature regularity of the spacetime.

\subsection{Surface gravities, temperatures,  and asymptotic behavior}
\label{sec2-2}
The construction makes the outer horizon $r_+$ a simple root and the inner horizon $r_-$ an effective triple root. Then the surface gravities satisfy
\begin{align}
\label{eq7}
{\kappa _ + } \! =\! \frac{1}{2}{\left. {\frac{{{\rm{d}}f\left( r \right)}}{{{\rm{d}}r}}} \right|_{r = {r_ + }}} \! \ne \! 0, \quad {\kappa _ - }\! =\!\ \frac{1}{2}{\left. {\frac{{{\rm{d}}f\left( r \right)}}{{{\rm{d}}r}}} \right|_{r = {r_ - }}}\! =\! 0,
\end{align}
which shows that the spacetime has a non-extremal outer horizon and a degenerate inner horizon. The corresponding horizon temperatures are given by
\begin{align}
\label{eq7+}
{T_ + } & \!=\! \frac{{{\kappa _ + }}}{{2\pi }}\! =\! \frac{{\left( {{r_ + } - {r_ - }} \right)}}{{8\left( {e - 1} \right)\pi M{r_ + }}}\left[ {\exp {{\left( {\frac{{{r_ - }}}{{{r_ + }}} - 1} \right)}^2} - 1} \right],
\nonumber\\
{T_ - } &= \frac{{{\kappa _ - }}}{{2\pi }} = 0.
\end{align}
The outer horizon has a nonzero temperature, whereas the formal quantity $T_-$ associated with the inner horizon vanishes. It is interesting to notice that with the evaporation of the black hole, $r_+=2M$ approaches $r_-$ such that the Hawking temperature $T_+$ goes to zero. As a result, the RBH would stop radiating and become a remnant with finite size at the Planck scale if we set $r_-$ at this order. In addition,  the structure near the inner horizon will play an important role in the analysis of the mass inflation problem below, where the double-null shell model and the Ori model will be used to test the late-time behavior of perturbations.

Next, we examine the asymptotic behavior of the spacetime. In the limit $r\to\infty$, one has $g \left(r \right)\to e-1$, which leads  to
\begin{align}
\label{eq8}
f\left(r\right) \to \frac{\left(r-r_-\right)\left(r-r_+\right)}{\left(r-r_-\right)\left(r-r_+\right)+2Mr}.
\end{align}
When expanding Eq.~(\ref{eq8}) at $ r\to\infty$, one has
\begin{align}
\label{eq9}
f\left(r\right) = 1-\frac{2M}{r}+\mathcal{O} \left(r^{-2} \right).
\end{align}
This indicates that the spacetime approaches the Schwa-rzschild form asymptotically, and the parameter $M$ can be identified as the ADM mass.

Furthermore, when $r\to0$, the exponential factor in Eq.~(\ref{eq6}) grows rapidly, so that both the numerator and the first term in the denominator are dominated by the same leading contribution. As a result,
\begin{align}
\label{eq10}
f\left(r\right)=1+\mathcal{O}\left({r^2}\right),\quad r\to0,
\end{align}
which implies that the metric function $f \left(r\right)$ remains finite at the center and does not exhibit the Schwarzschild-type divergence. These asymptotic and near center behaviors will be essential for the curvature analysis in the next subsection, where the regularity of the spacetime and the global curvature bound will be examined in detail.

\subsection{Curvature regularity and global sub-Planckianity}
\label{sec2-3}
We now examine the curvature invariants to determine the regularity of the spacetime and whether its curvature scale can remain everywhere below the Planck scale. For this purpose, we consider the Ricci scalar $R$, the quadratic Ricci invariant $R_{\mu\nu}R^{\mu\nu}$, and the Kretschmann scalar $K=R_{\mu\nu\rho\sigma}R^{\mu\nu\rho\sigma}$. Their explicit expressions are presented in~\ref{appA}. In the limit $r\to 0$, one has
\begin{align}
\label{eq11}
R=0,\quad R_{\mu\nu}R^{\mu\nu}=0,\quad K=0.
\end{align}
This confirms that the spacetime is free of a curvature singularity at the center. Moreover, this result are independent of $M$ and $r_\pm$. Combined with the near center behavior in Eq.~(\ref{eq10}), it follows that the spacetime approaches a Minkowski core as $r\to 0$. It is noted that the present analysis concerns curvature invariants constructed algebraically from the Riemann tensor and the metric, while invariants involving covariant derivatives of the curvature are not considered.
\begin{figure}[htbp]
\centering
\subfigure[]{
\begin{minipage}{0.47\textwidth}
\includegraphics[width=\textwidth]{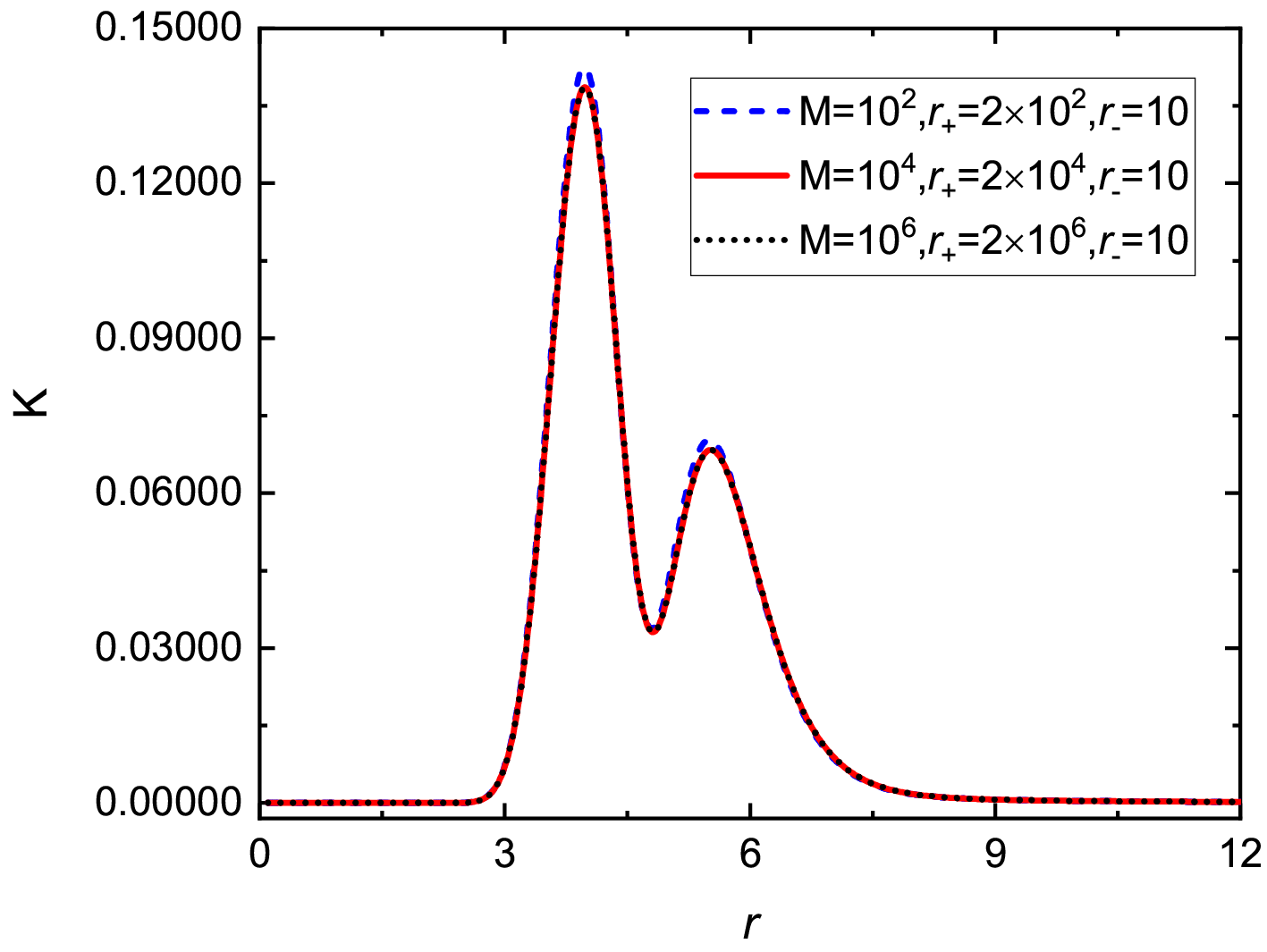}
\label{fig1-1}
\end{minipage}
}
\subfigure[]{
\begin{minipage}{0.47\textwidth}
\includegraphics[width=\textwidth]{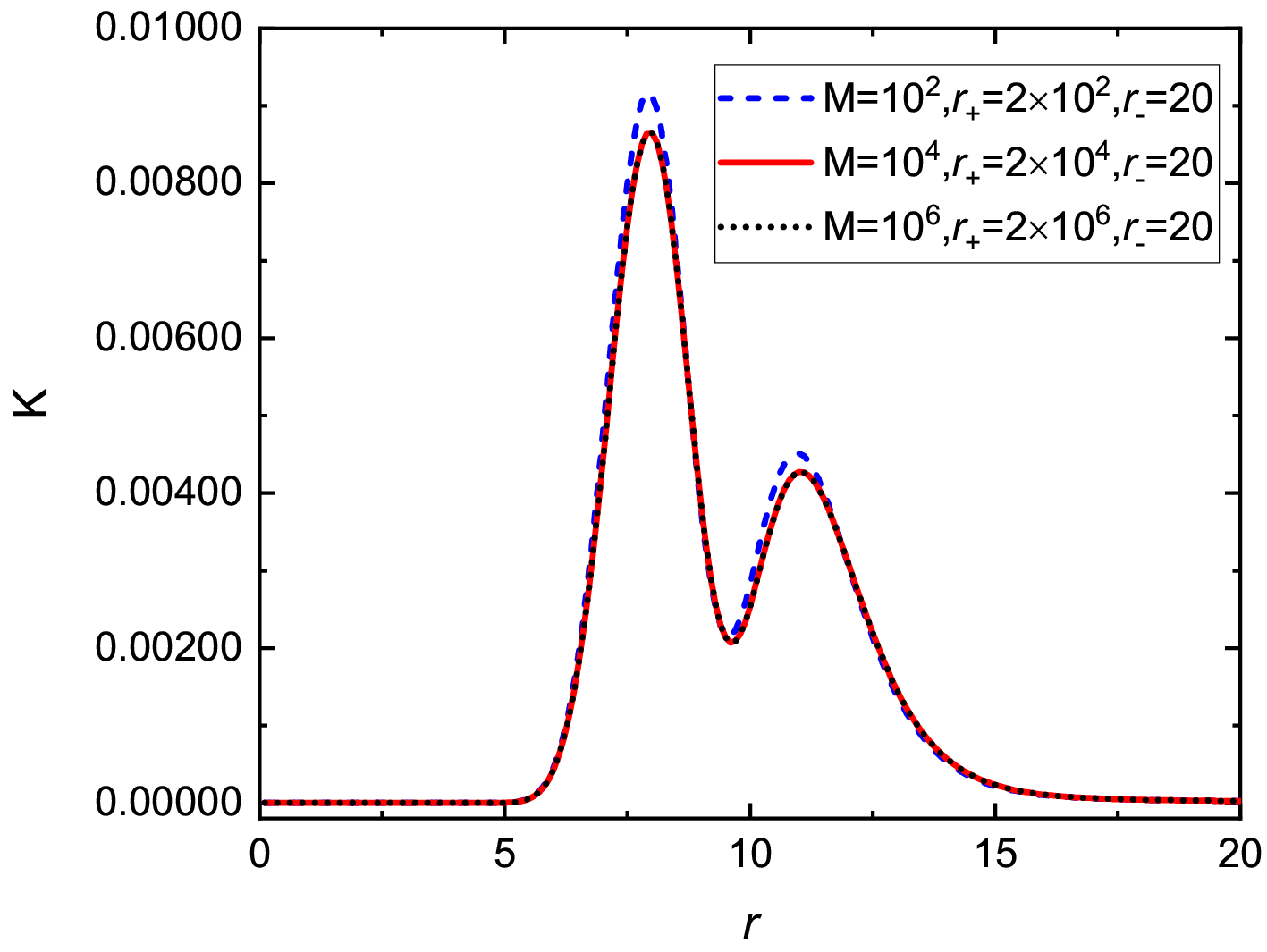}
\label{fig1-2}
\end{minipage}
}
\caption{The Kretschmann scalar $K$ as a function of $r$ for different black hole masses. (a) Fixed $r_-=10$. (b) Fixed $r_-=20$.}
\label{fig1}
\end{figure}

Although central regularity is essential, it is not sufficient for our purpose. For the present spacetime, the more relevant question is whether the curvature can remain below the Planck scale everywhere, especially for large black holes. To address this question, we take $r_+=2M$ and treat $r_-$ as an independent internal scale. We shall be mainly interested in the parameter range $0<r_-\ll r_+$, which corresponds to a macroscopic outer horizon and a much smaller internal scale. Under these conditions, the behavior of the Kretschmann scalar is shown in Fig.~\ref{fig1}.

In both panels of Fig.~\ref{fig1}, the blue dashed curve for $M=10^2$ lies visibly above the red solid curve for $M=10^4$ and the black dotted curve for $M=10^6$, while the latter two are nearly indistinguishable. The comparatively modest values of $M$ are chosen deliberately to resolve the approach to the large-mass regime. For astrophysical black hole masses expressed in Planck units, the curvature profiles would already have converged to essentially the same large-mass behavior. Thus, the Kretschmann scalar shows an appreciable dependence on $M$ only when the mass is relatively small, while this dependence becomes strongly suppressed as $M$ increases. Moreover, comparing Fig.~\ref{fig1-1} with Fig.~\ref{fig1-2}, one finds that the overall curvature scale is significantly reduced when $r_-$ is increased. This suggests that, in the large-mass regime, the curvature scale is governed mainly by $r_-$ rather than by $M$. This behavior can be understood by introducing the dimensionless variables
\begin{align}
\label{eq12}
u=\frac{r}{r_-},\quad \varepsilon=\frac{r_-}{2M},
\end{align}
with $r_+=2M$. Then, the metric function~(\ref {eq6}) can be rewritten as
\begin{align}
\label{eq13}
\tilde f\left( u \right) = \frac{{g\left( u \right)\left( {u - 1} \right)\left( {\varepsilon u - 1} \right)}}{{g\left( u \right)\left( {u - 1} \right)\left( {\varepsilon u - 1} \right) + \left( {e - 1} \right)u}}.
\end{align}

According to  Eq.~(\ref{eq13}), the Kretschmann scalar in the spherically symmetric static spacetime can be expressed as
\begin{align}
\label{eq14}
K\left( u, \varepsilon \right) & = {\left( {\frac{1}{{r_ - ^2}}{{\tilde f}_{uu}}} \right)^2} + \frac{4}{{{r^2}}}{\left( {\frac{1}{{{r_ - }}}{{\tilde f}_u}} \right)^2} + \frac{{4{{\left( {1 - \tilde f} \right)}^2}}}{{{r^4}}}
\nonumber \\
& = \frac{1}{{r_ - ^4}}\left[ {\tilde f_{uu}^2 + \frac{{4\tilde f_u^2}}{{{u^2}}} + \frac{{4\left( {1 - \tilde f} \right)^2}}{{{u^4}}}} \right],
\end{align}
where ${{\tilde f}_{u}}={\rm d}{{\tilde f}}\left(u \right)/{\rm d}{u}$ and ${{\tilde f}_{uu}}={\rm d^2}{{\tilde f}}\left(u \right)/{\rm d}{u^2}$. In the large-mass regime one has $\varepsilon \ll 1$, and Eq.~(\ref{eq14}) can be expanded as
\begin{align}
\label{eq15}
K\left( u, \varepsilon \right) &= \frac{1}{{r_ - ^4}}A\left( {u,\varepsilon } \right) = \frac{1}{{r_ - ^4}}{A_0}\left( u \right) + \frac{\varepsilon }{{r_ - ^4}}{A_1}\left( u \right) 
\nonumber\\
&+ \frac{{{\varepsilon ^2}}}{{r_ - ^4}}{A_2}\left( u \right) +  \cdots ,
\end{align}
where the functions $A_i\left(u\right)$ depend only on the dimensionless variable $u$. This expansion makes explicit that the leading contribution to the Kretschmann scalar is independent of $M$, while the residual mass dependence enters only through subleading corrections of order $\varepsilon\propto M^{-1}$\footnote{In the large-mass regime, this further reduces to the Schwarzschild scaling ${T_ + } = 1/8\pi M$, while the formal quantity associated with the inner horizon satisfies $T_-=0$.}. In the large-mass regime, the curvature distribution becomes only weakly sensitive to $M$, and its overall scale is set predominantly by $r_-^{-4}$. This analytical result is in agreement with Fig.~\ref{fig1}, where the curves for $M=10^4$ and $M=10^6$ are nearly indistinguishable.

Having shown how the curvature distribution behaves in the large-mass regime, we next examine whether the spacetime can remain everywhere below the Planck scale. For this purpose, it is sufficient to examine the global maximum of the Kretschmann scalar, denoted by $K_{\max}$, and the radius $r_{\max}$ at which it is attained. Since the spacetime is asymptotically Schwarzschild, one has $K\left(r\right)\to 0$ as $r\to \infty$, while Eq.~(\ref{eq11}) shows that $K\left(r\right)\to0$ as $r\to 0$. Therefore, the global maximum of $K\left(r\right)$ must be attained at a finite radius $r=r_{\max}$. The condition $K_{\max}<1$ then implies $K\left(r\right)<1$ for all $r\in[0,\infty)$, namely, that the spacetime is everywhere sub-Planckian. To determine this global maximum, it is convenient to work with the dimensionless variables introduced above. In terms of $K\left(u,\varepsilon\right)$, the position of the global maximum $u$ is determined by ${{{\rm{d}}K\left( {u,\varepsilon } \right)} \mathord{\left/ {\vphantom {{{\rm{d}}K\left( {u,\varepsilon } \right)} {{\rm{d}}u}}} \right. \kern-\nulldelimiterspace} {{\rm{d}}u}} = 0$.  According to Eq.~(\ref{eq15}), this condition is equivalent to
\begin{align}
\label{eq15+}
A'\left(u,\varepsilon\right)=A_0'\left(u\right)+\varepsilon A_1'\left(u\right)+\varepsilon^2 A_2'\left(u\right)+\cdots=0,
\end{align}
where the prime denotes differentiation with respect to $u$. At leading order, one sets $\varepsilon=0$, and the corresponding maximum position $u_0$ is determined by ${{A'}_0}\left( {{u_0}} \right) = 0$ and ${{A''}_0}\left( {{u_0}} \right) < 0$, which leads to
\begin{align}
\label{eq16+}
{K_{\max \left( 0 \right)}} = \frac{1}{{r_ - ^4}}{A_0}\left( {{u_0}} \right).
\end{align}
Furthermore, including higher-order corrections, the global maximum of the Kretschmann scalar takes the form
\begin{align}
\label{eq16}
{K_{\max }} = \frac{1}{{r_ - ^4}}\left[ {{A_0}\left( {{u_0}} \right) + \varepsilon {A_1}\left( {{u_0}} \right) +{\mathcal{O}}\left( {\varepsilon^2} \right)} \right],
\end{align}
where ${{A_0}\left( {{u_0}} \right)}$ and ${{A_1}\left( {{u_0}} \right)}$ are constants, and the detailed derivation is presented in~\ref{appB}. Eq.~(\ref{eq16}) shows that the dominant term in $K_{\max}$ is controlled by $r_-$ in the large-mass limit, while the dependence on $M$ appears only as a higher-order correction. This explains why an appropriate choice of $r_-$ is sufficient to keep the maximal curvature below the Planck scale.

\begin{figure*}[htbp]
\centering
\subfigure[]{
\begin{minipage}{0.45\textwidth}
\includegraphics[width=\textwidth]{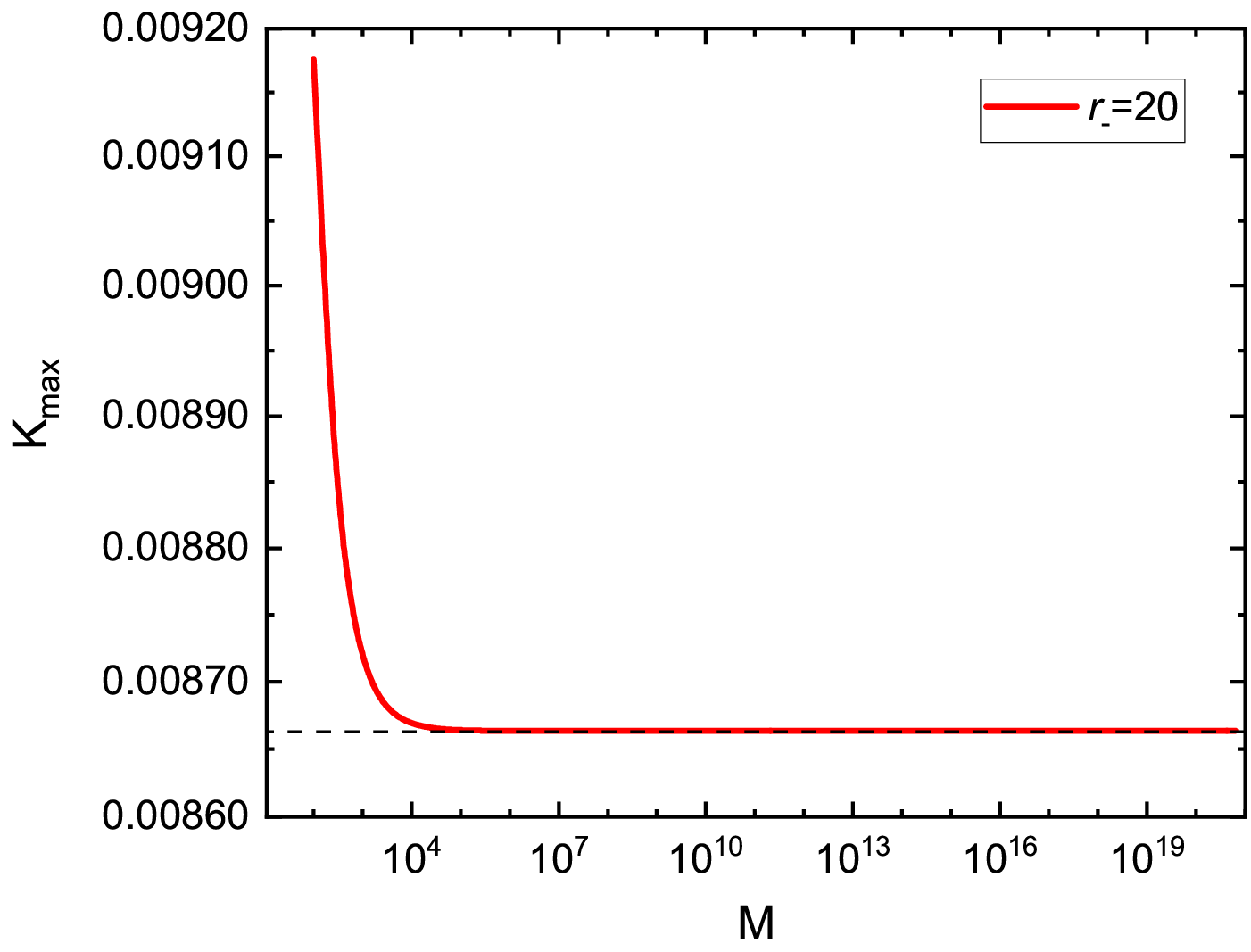}
\label{fig2-1}
\end{minipage}
}
\subfigure[]{
\begin{minipage}{0.45\textwidth}
\includegraphics[width=\textwidth]{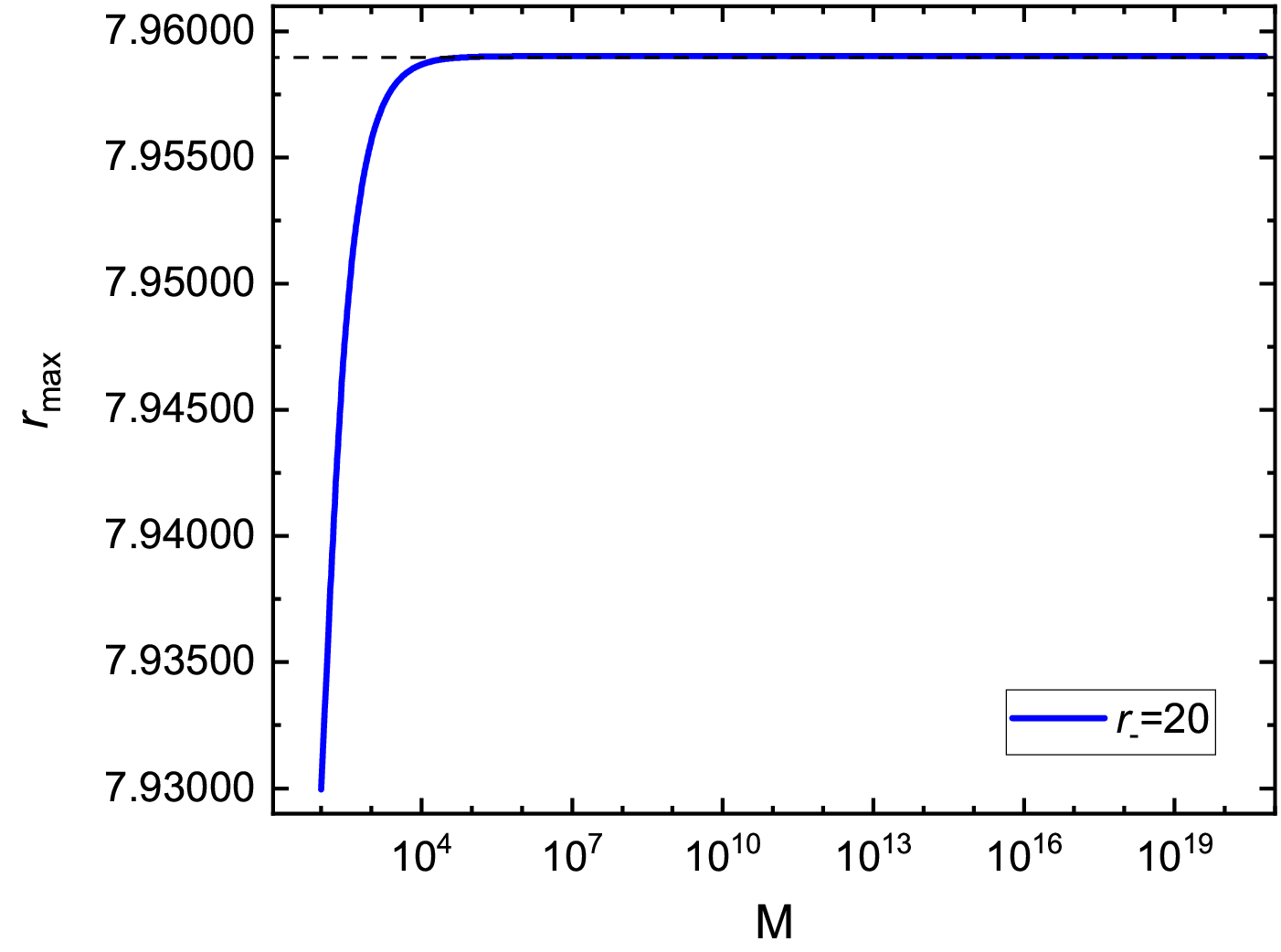}
\label{fig2-2}
\end{minipage}
}
\subfigure[]{
\begin{minipage}{0.45\textwidth}
\includegraphics[width=\textwidth]{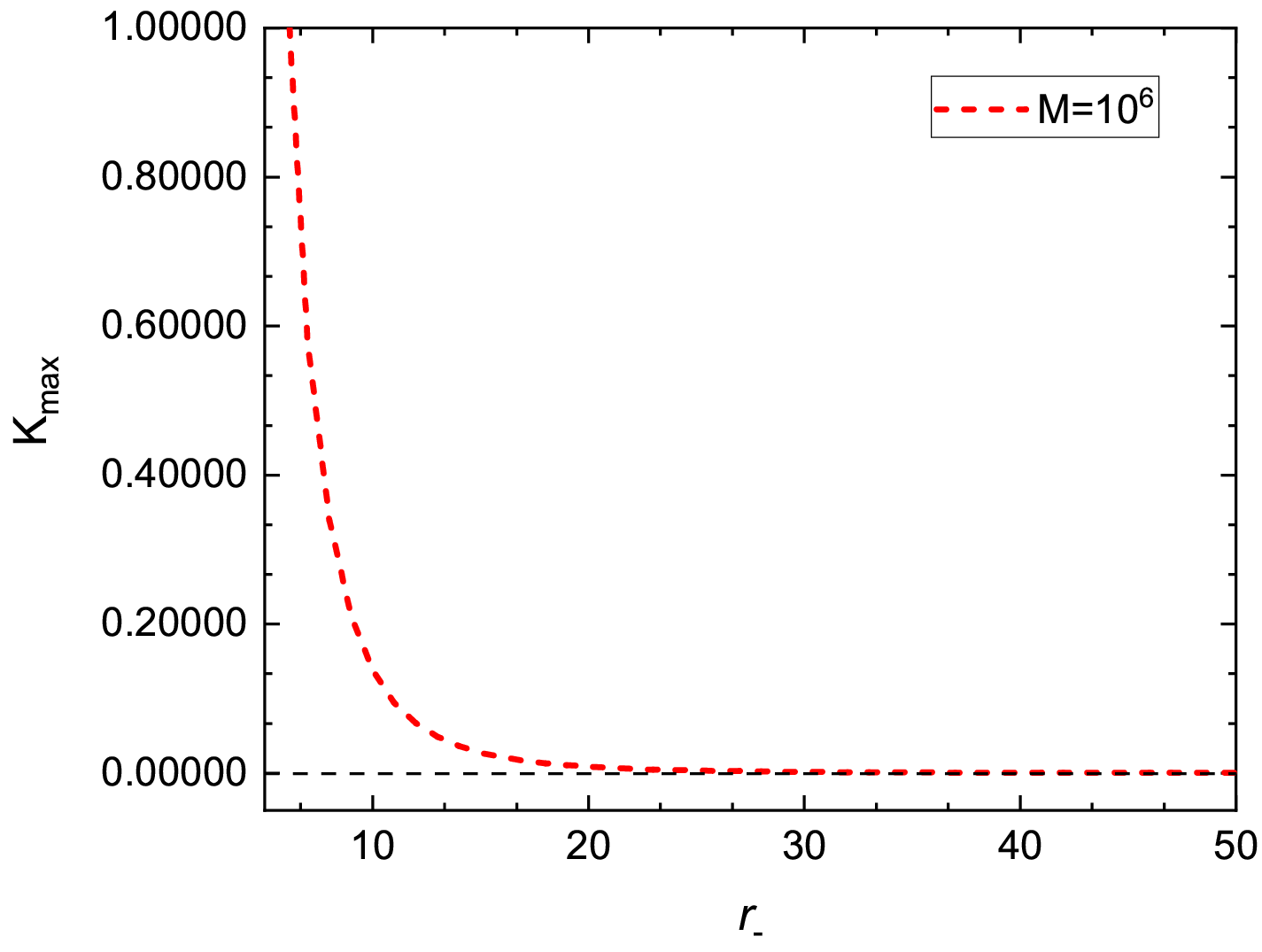}
\label{fig2-3}
\end{minipage}
}
\subfigure[]{
\begin{minipage}{0.45\textwidth}
\includegraphics[width=\textwidth]{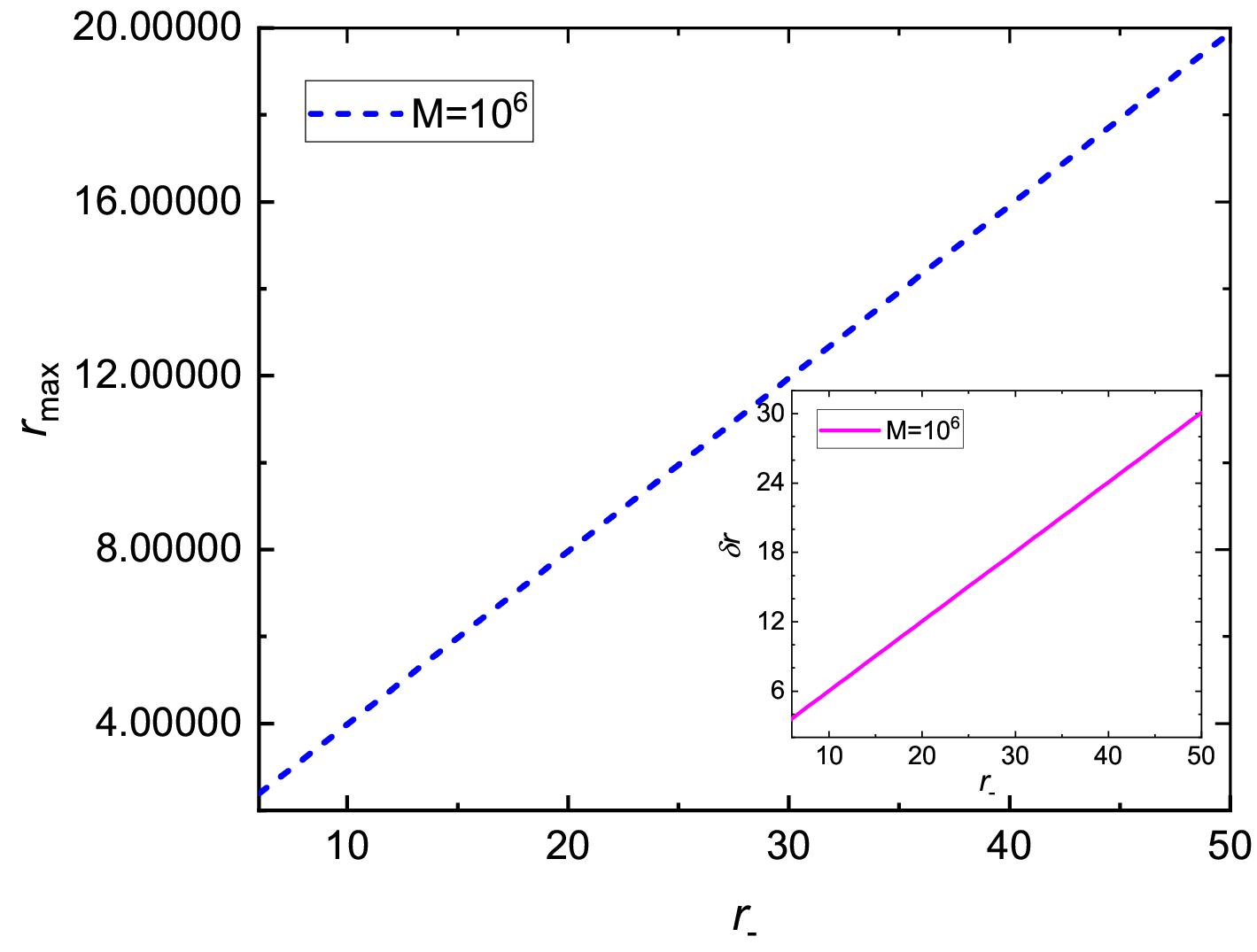}
\label{fig2-4}
\end{minipage}
}
\caption{Numerical behavior of the global maximum $K_{\max}$ of the Kretschmann scalar and its location $r_{\max}$. (a) and (b) show the dependence on the mass $M$ for fixed $r_-=20$. (a) $K_{\max}$ as a function of $M$. In the large-mass regime, $K_{\max}\approx0.00866$. (b) The corresponding location $r_{\max}$ as a function of $M$. In the large-mass regime, $r_{\max}\approx7.95902$. (c) and (d) show the dependence on the inner horizon radius $r_-$ for fixed $M=10^6$ and $r_+=2M$. (c) $K_{\max}$ as a function of $r_-$. (d) The corresponding location $r_{\max}$ as a function of $r_-$. The inset shows $\delta r= r_- - r_{\max}$.}
\label{fig2}
\end{figure*}
In addition to the analytical arguments given above, Fig.~\ref{fig2} provides a direct numerical confirmation of the behavior of the global curvature bound. Figs.~\ref{fig2-1} and~\ref{fig2-2} show that, for fixed $r_-=20$, both $K_{\max}$ and the corresponding location $r_{\max}$ rapidly approach constant values as $M$ increases. This behavior is consistent with Eqs.~(\ref{eq15}) and (\ref{eq16}), and confirms that the global curvature scale becomes essentially insensitive to $M$ in the large-mass regime. Figs.~\ref{fig2-3} and~\ref{fig2-4} clarify the role of the inner horizon. For fixed $M=10^6$ and $r_+=2M$, Fig.~\ref{fig2-3} shows that $K_{\max}$ decreases rapidly as $r_-$ increases, in agreement with the analytical result that the dominant curvature scale is controlled by $r_-^{-4}$. From Fig.~\ref{fig2-4}, one can see that the position $r_{\max}$ of the global maximum increases with $r_-$. The inset displays the difference $\delta r = {r_-} - r_{\max}$, which remains positive throughout the parameter range considered. For the cases considered, the global curvature maximum lies inside the inner horizon. These results confirm that the overall curvature scale is governed predominantly by the inner horizon $r_-$ rather than by the ADM mass $M$. Consequently, by choosing $r_-$ appropriately, one can ensure that the Kretschmann scalar remains below the Planck scale everywhere in the spacetime. In the large-mass regime considered here, the numerical analysis gives the sufficient bound $r_-\gtrsim 6.10174$ for keeping $K\left(r\right)<1$ for $0 \leq r < \infty$. A more quantitative discussion of the corresponding lower bound on $r_-$ is deferred to~\ref{appC}.

\subsection{Null energy condition}
\label{sec2-4}
The analysis of energy conditions provides a simple diagnostic of the effective stress-energy tensor reconstructed from the geometry. Here we focus on the null energy condition (NEC), defined by $T_{\mu\nu}k^\mu k^\nu \ge 0$ with any null vector $k^\mu$. In the static regions, the effective energy-momentum tensor can be written in the form $T^\mu_{ \nu} = \mathrm{diag}\left(-\rho, p_r, p_t, p_t \right)$, where the energy density $\rho$, radial pressure $p_r$, and tangential pressure $p_t$ are given by
\begin{align}
\label{eq16++}
\rho \!=\! \frac{1\! -\! f\! - \!r f'}{8\pi r^2}, p_r\! =\! \frac{f\! -\! 1 \!+\! r f'}{8\pi r^2}, p_t\! = \!\frac{f''}{16\pi} +\frac{f'}{8\pi r}.
\end{align}
It follows immediately that $p_r=-\rho$, so that the NEC along radial null directions is identically saturated,
\begin{align}
\label{eq17+}
R_{\rm{NEC}}=\rho + p_r = 0.
\end{align}
The remaining, nontrivial part of the NEC is associated with null directions involving the angular sector. In the regions $0<r<r_-$ and $r_+<r<\infty$, where $t$ remains timelike, the corresponding null contraction reduces to the combination $\rho+p_t$. One yields
\begin{align}
\label{eq18+}
A_{\rm{NEC}}=\rho + p_t = \frac{1}{8\pi}\left(\frac{1-f}{r^2}+\frac{f''}{2}\right).
\end{align}
In the trapped region $r_-<r<r_+$, the coordinate $r$ becomes timelike.  In this region, the effective energy density should be defined with respect to the timelike $r$ direction, namely $\rho_{\rm trap}=-T^r_{\ r}=-p_r $. The corresponding nontrivial null contraction is then given by $\rho_{\rm trap}+p_t=-p_r+p_t$, and one finds
\begin{align}
\label{eq18++}
A_{\rm{NEC}}=\rho_{\rm trap}+p_t=\frac{1}{8\pi}\left(\frac{1-f}{r^2}+\frac{f''}{2}\right).
\end{align}
Thus, the NEC along radial null directions is identically saturated, while its remaining nontrivial part throughout the spacetime is governed by Eqs.~(\ref{eq18+})-(\ref{eq18++}).

\begin{figure}[htbp]
\centering 
\includegraphics[width=.45\textwidth,origin=c,angle=0]{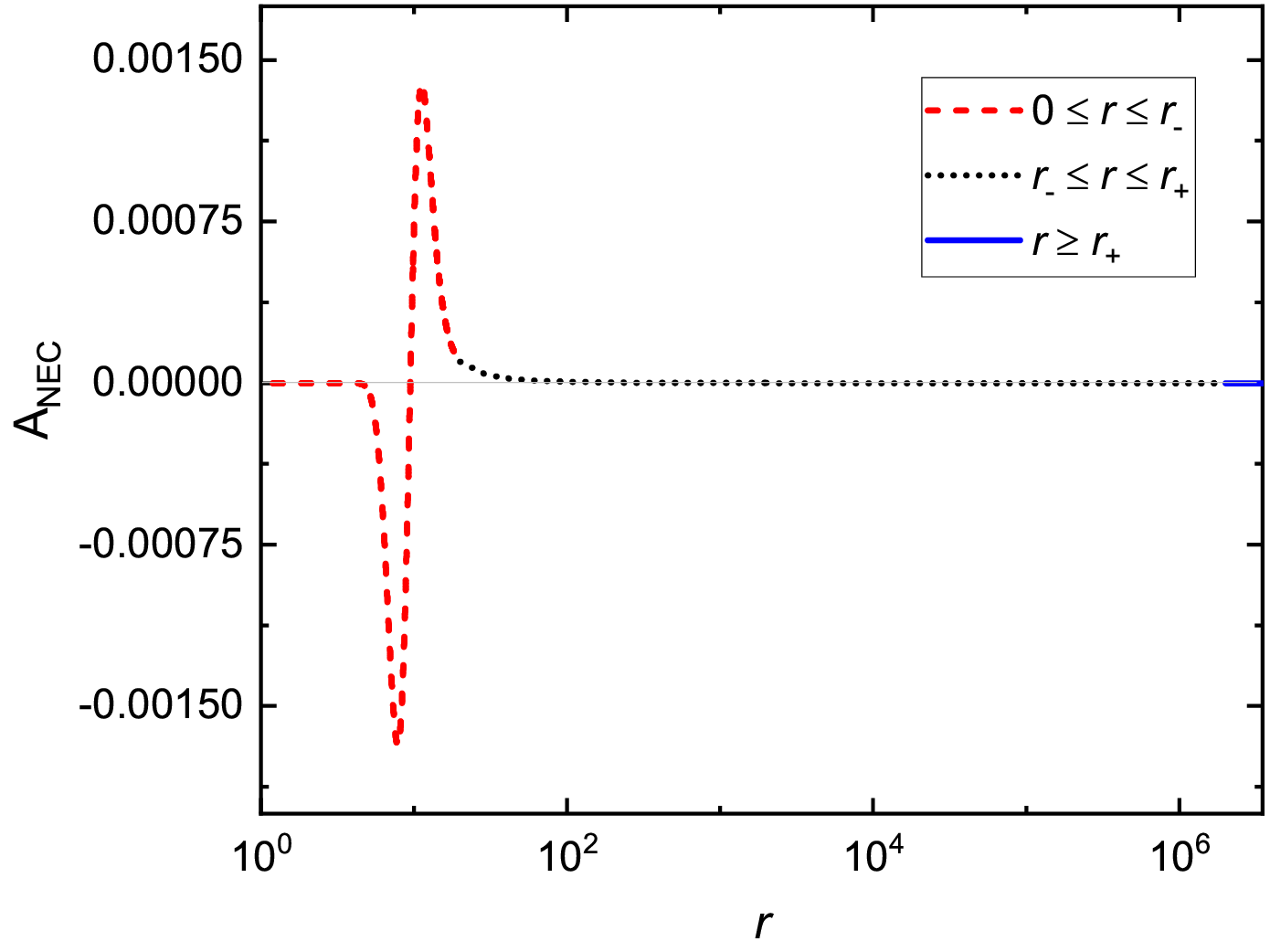}
\caption{\label{fig5} $A_{\rm{NEC}}$ as a function of $r$ for fixed $M=10^6$, $r_+=2M$, and $r_-=20$.}
\end{figure}
In order to understand the behavior of the NEC, we examine its limiting behavior at the regular center, the inner horizon, and in the asymptotic region. At the regular center $r\to0$, the spacetime approaches a Minkowski core and the NEC is saturated. At the inner horizon $r=r_-$, the metric function satisfies $f(r_-)=f'(r_-)=f''(r_-)=0$, from which Eq.~(\ref{eq18+}) gives ${\left. {{A_{{\rm{NEC}}}}} \right|_{r = {r_ - }}} = {1 \mathord{\left/ {\vphantom {1 {8\pi r_ - ^2}}} \right. \kern-\nulldelimiterspace} {8\pi r_ - ^2}} > 0$, indicating that the NEC is strictly satisfied at the inner horizon. In the asymptotic region, $A_{\rm NEC}$ approaches zero. As shown in Fig.~\ref{fig5}, for $M=10^6$, $r_+=2M$, and $r_-=20$, the most significant NEC violation occurs within the finite interval $2.53923\lesssim r\lesssim9.46275$ below the inner horizon. Outside this region, the violation is strongly suppressed and becomes negligible at large distances.

Having established the global geometric role of the inner-horizon scale $r_-$, we next turn to the dynamics near the degenerate inner horizon.

\section{Tests of the exponential mass inflation mechanism}
\label{sec3}
The relevant physical question is not merely whether the inner horizon has vanishing surface gravity, but how the order of the zero of the metric function modifies the accumulation of blueshift along outgoing trajectories. In the present geometry, the cubic behavior of $f\left(r\right)$ makes outgoing null rays approach the inner horizon algebraically, so that the geometric blueshift factor grows as a power law rather than exponentially. The following three analyses examine different consequences of this mechanism. The scalar-field calculation establishes the local near-horizon scaling, the DTR model tests its effect on the nonlinear shell-crossing term, and the Ori model examines the resulting mass evolution in the presence of a continuous Price-law ingoing flux.

\subsection{Massless scalar test field}
\label{sec3-1}
We begin with a massless scalar test field on the fixed background, which satisfies $\square \Phi=0$. For the static spherically symmetric metric~(\ref{eq1}), the field equation can be written as
\begin{align}
\label{eq1a1}
-\frac{1}{f}\frac{\partial^2\Phi}{\partial t^2}
+\frac{1}{r^2}\frac{\partial}{\partial r}
\left(
r^2 f\frac{\partial\Phi}{\partial r}
\right)
+\frac{1}{r^2}\nabla^2_{\Omega}\Phi=0 ,
\end{align}
where $\nabla^2_{\Omega}$ is the Laplace operator on the unit two-sphere. Furthermore, the scalar field can be decomposed in terms of spherical harmonics,
\begin{align}
\label{eq1a2}
\Phi\left(t,r,\theta,\varphi\right)
= \sum_{l m} \frac{e^{-i\omega t}\psi_ l \left(r\right)}{r} Y_{l m}(\theta,\varphi),
\end{align}
with $\nabla^2_{\Omega}Y_{l m} =-l \left(l+1 \right)Y_{l m}$. In order to cast the radial part of Eq.~(\ref{eq1a1}), we introduce the tortoise coordinate ${{\rm d}r_*}/{{\rm d}r}=1/{f\left(r\right)}$. Then, the scalar perturbation equation is
\begin{align}
\label{eq1a5}
\frac{{\rm d}^2\psi_\ell}{{\rm d}r_*^2}
+\left[\omega^2-V_\ell\left(r\right)\right]\psi_\ell=0,
\end{align}
where the effective potential is $V_\ell\left(r\right) = f [l\left(l+1 \right)/r^2+f'/r ]$. Since the effective potential  is determined by $f\left(r\right)$ and $f'\left(r\right)$, its near-horizon behavior is fixed by the order of zero of $f\left(r\right)$. We therefore expand the metric function around the inner horizon. For this purpose, we define $\delta=r-r_-$. From Eq.~(\ref{eq6}), the metric function near the inner horizon behaves as follows
\begin{align}
\label{eq1a6}
f\left(r\right) \! = \!-\alpha_-\delta^3 \!+\!\mathcal{O}\left(\delta^4\right),
f'\left(r\right) \!=\! -3\alpha_-\delta^2\!+\!\mathcal{O}\left(\delta^3\right),
\end{align}
where we consider the approach to the inner horizon from the trapped region $r_-<r<r_+$, and $\alpha_-=-f'''\left(r_-\right)/6= \left ( r_+-r_-\right )/\left[2\left ( e-1\right ) M r^3_-\right]>0$. With the expansions~(\ref{eq1a6}), the leading near-horizon behavior of the scalar effective potential is given by
\begin{align}
\label{eq1a8}
V_\ell\left(r \right) = -\alpha_-
\frac{l \left(l+1 \right)}{r_-^2}
\delta^3+\mathcal{O}\left( \delta ^4\right ), \quad l\geq 1,
\end{align}
while the leading term for the $\ell=0$ mode starts at order $\delta^5$. The cubic suppression of the potential reflects the same near-horizon structure that controls the blueshift behavior of outgoing null rays.

The cubic zero of the metric function also controls the late-time approach of outgoing null rays to the inner horizon. To describe this behavior, we use the ingoing Eddington-Finkelstein coordinate $v=t+r_*$, in which radial outgoing null rays satisfy ${\rm d}r/{\rm d}v=f\left ( r\right )/2$. Using the near-horizon expansion in Eq.~(\ref{eq1a6}) and defining
$\delta \left(v\right)= r \left(v\right)-r_-$, one has
\begin{align}
\label{eq1a9}
\frac{{\rm d}\delta}{{\rm d}v} = -\frac{\alpha_-}{2}\delta^3.
\end{align}
After integration, one obtains $\delta\left(v\right) \simeq \alpha_-^{-1/2}\left(v-v_* \right)^{-1/2}$, where $v_*$ is an integration constant. Substituting $\delta\left(v\right)$ into the near-horizon expansion $\left|f\right|\simeq \alpha_-\delta^3$, one gets
\begin{align}
\label{eq1a10}
\left|f\left(r\left(v\right)\right)\right|
& \simeq \alpha_-^{-1/2}\left(v-v_*\right)^{-3/2}
\nonumber\\
& =\left[
\frac{2\left(e-1\right)Mr_-^3}{r_+-r_-}
\right]^{1/2}\left(v-v_*\right)^{-3/2}.
\end{align}
The above equation shows that, along outgoing null rays, the metric function approaches zero in magnitude only as a power law. Consequently, the amplification factor associated with the inner-horizon blueshift behaves as $\left|f\left(r\left(v\right)\right)\right|^{-1} \simeq \alpha_-^{1/2}\left(v-v_*\right)^{3/2}$. For comparison, a generic nondegenerate inner horizon is characterized by a simple zero of the metric function $f_{\rm nd}\left(r\right) \simeq 2\kappa_{\rm nd}\left(r-r_-\right)$ and $\kappa_{\rm nd}<0$. In this case, an outgoing null ray approaches the inner horizon exponentially $r\left(v\right)-r_- \propto {\rm exp} \left(\kappa_{\rm nd}v\right)$, and the corresponding amplification factor grows as $\left|f_{\rm nd}\left(r\left(v\right)\right)\right|^{-1} \propto {\rm exp} \left(\left|\kappa_{\rm nd}\right|v\right)$. The cubic degeneracy replaces the standard exponential blueshift
amplification with a power-law growth. The scaling derived above provides the common near-horizon input for the double-null shell and Ori models considered below.

\subsection{Test based on the double-null shell model}
\label{sec3-2}

Having established the local near-horizon scaling in the preceding subsection, we now apply the double-null shell model proposed by Dray, 't Hooft, and Redmount (DTR)~\cite{Dray:1985yt,Redmount:1985aoy}. In the present geometry, the model-specific information enters through the response of $m\left(r,M\right)$ to a small change in $M$ at fixed $r_-$. We therefore examine how this response combines with the near-horizon scaling in the nonlinear shell-crossing contribution to the Misner-Sharp mass. For the metric in Eq.~(\ref{eq1}), the Misner-Sharp mass is $m\left(r\right)=r\left[1-f\left(r\right)\right]/2$. In the DTR model, an ingoing null shell and an outgoing null shell intersect inside the black hole and divide the spacetime into four regions. The corresponding Misner-Sharp masses are related by the DTR matching  condition~\cite{CarballoRubio2018},
\begin{align}
\label{eq19}
{m_{\rm{f}}}\left( {{r_ \times }} \right) & = {m_{\rm{i}}}\left( {{r_ \times }} \right) + {m_{{\rm{in}}}}\left( {{r_ \times }} \right) + {m_{{\rm{out}}}}\left( {{r_ \times }} \right)
\nonumber\\
& - \frac{{2{m_{{\rm{in}}}}\left( {{r_ \times }} \right){m_{{\rm{out}}}}\left( {{r_ \times }} \right)}}{{{r_\times}{f_{\rm{i}}}\left( {{r_ \times }} \right)}},
\end{align}
where $r_ \times$ denotes the shell-crossing radius,  $m_{\rm f}$ and $m_{\rm i}$ are the Misner-Sharp mass functions in the future and initial regions, respectively, while $m_{\rm in}$ and $m_{\rm out}$ denote the variations of the Misner-Sharp mass induced by the ingoing and outgoing shells, and are evaluated to leading order as the response of $m\left(r,M\right)$ to small shifts of the mass parameter across the ingoing and outgoing shells. The last term in Eq.~(\ref{eq19}) is the nonlinear contribution responsible for the standard mass inflation behavior. Its late-time behavior is determined by the near-horizon scalings of $f_{\rm i}$, $m_{\rm in}$, and $m_{\rm out}$.

As in the curvature analysis, we set $r_+=2M$ and treat $r_-$ as an independent internal scale. The shell-induced variations are evaluated by comparing neighboring geometries with slightly different ADM masses at fixed $r_-$. Using Eqs.~(\ref{eq2}) and~(\ref{eq6}), the Misner-Sharp mass can be written as
\begin{align}
\label{eq19++}
m\left(r,M\right) \!= \!\frac{\!\left(e-1\right)Mr^2} {g\left(r\right)\left(r-r_-\right)\!\left(r-2M\right)
\!+\!2\!\left(e-1\right)Mr}.
\end{align}
Since $g\left(r\right)$ is independent of $M$ when $r_-$ is fixed, differentiation with respect to the ADM mass gives
\begin{align}
\label{eq19+}
\left(\frac{\partial m}{\partial M}
\right)_{r_-}& \!= \!\frac{\left(e-1\right)g\left(r\right)r^3\left(r-r_-\right)}{\left[g\left(r\right)\left(r-r_-\right)\left(r-2M\right)+2\left(e-1\right)Mr
\right]^2}
\nonumber\\
&\!=\!\frac{\left(r-r_-\right)^3}{4\left(e-1\right)M^2r_-}+\mathcal{O}\left[\left(r-r_-\right)^4\right],
\end{align}
where we have used $g\left(r\right)=\left(r-r_-\right)^2/r_-^2+\mathcal{O}\left[\left(r-r_-\right)^3\right]$ near the inner horizon. Eq.~(\ref{eq19+}) explicitly shows that the first nonvanishing mass response is cubic in $r-r_-$. For a more general family with $r_-=r_-\left(M\right)$, the total response also contains the contribution
$(\partial m/\partial r_-)$ $({\rm d}{r_-}/{\rm d}M)$, which can begin at quadratic order in $r-r_-$~\cite{CarballoRubio2022}. This lower-order contribution is absent in the fixed $r_-$ parametrization considered here. To leading order in the shell amplitudes and in $r_\times-r_-$, one obtains
\begin{subequations}
\label{eq20}
\begin{align}
&{m_{{\rm{in}}}}\left( {{r_ \times },M,{M_{{\rm{in}}}}} \right)=
{\mathcal{D}}\left( {{r_ - },M} \right)
{\left( {{r_ \times } - {r_ - }} \right)^3}
{M_{{\rm{in}}}},
\\
&{m_{{\rm{out}}}}\left( {{r_ \times },M,{M_{{\rm{out}}}}} \right) \!=\!
{\mathcal{D}}\left( {{r_ - },M} \right)
{\left( {{r_ \times } - {r_ - }} \right)^3}
{M_{{\rm{out}}}},
\end{align}
\end{subequations}
where ${\mathcal{D}}\left( {{r_ - },M} \right) = {1 \mathord{\left/ {\vphantom {1 {4\left( {e - 1} \right){M^2}{r_ - }}}} \right. \kern-\nulldelimiterspace} {4\left( {e - 1} \right){M^2}{r_ - }}}$. The shell-induced mass perturbations therefore do not vanish identically, but they are cubically suppressed as the shell-crossing point approaches the degenerate inner horizon.

The crossing point follows the outgoing null shell and therefore obeys the same near-horizon scaling derived in Sec.~\ref{sec3-1}. Defining
$\delta_\times \left(v\right)=r_\times \left(v\right)-r_-$, one has
\begin{subequations}
\label{eq21}
\begin{align}
\delta_\times\left(v\right)
&\simeq \alpha_-^{-1/2} \left(v-v_*\right)^{-1/2},
\\
f_{\rm i}\left(r_\times(v)\right)
&\simeq -\alpha_-^{-1/2}
\left(v-v_*\right)^{-3/2}.
\end{align}
\end{subequations}
Substituting Eq.~(\ref{eq21}) into Eq.~(\ref{eq20}), one obtains
\begin{subequations}
\label{eq22}
\begin{align}
&{m_{{\rm{out}}}}\!\left( {{r_ \times }\left( v \right),M,{M_{{\rm{out}}}}} \right) \!= \!{\mathcal{H}}\!\left( {{r_ - },M} \right){\left( {v - {v_*}} \right)^{ - \frac{3}{2}}}{M_{{\rm{out}}}},
\\
&{m_{{\rm{in}}}}\!\left( {{r_ \times }\left( v \right),M,{M_{{\rm{in}}}}\!\left( v \right)} \right) \! =\! {\mathcal{H}}\!\left( {{r_ - },M} \right){\left( {v - {v_*}} \right)^{ - \frac{3}{2}}}{M_{{\rm{in}}}}\!\left( v \right),
\end{align}
\end{subequations}
where ${\mathcal{H}}\left( {{r_ - },M} \right)={\left[ {\frac{{2(e - 1)Mr_ - ^3}}{{{r_ + } - {r_ - }}}} \right]^{3/2}}{\mathcal{D}}\left( {{r_ - },M} \right)$. Since the crossing point approaches the inner horizon from the trapped
region, $f_{\rm i}<0$, and the nonlinear term in Eq.~(\ref{eq19}) is
positive. Using Eqs.~(\ref{eq21}) and~(\ref{eq22}), its late-time
behavior is
\begin{align}
\label{eq23}
-\frac{2m_{\rm in}m_{\rm out}}{
r_\times f_{\rm i}} \simeq
\frac{2\mathcal{H}^2\alpha_-^{1/2}}{
r_-}M_{\rm in}(v)M_{\rm out}
\left(v-v_*\right)^{-3/2}.
\end{align}
The nonlinear term in Eq.~(\ref{eq23}) decays as $\left(v-v_*\right)^{-3/2}$ rather than growing exponentially. If the shell amplitudes remain finite, or more generally decay sufficiently rapidly,
both the linear and nonlinear shell-induced contributions vanish at late times. The DTR relation then takes the asymptotic form
\begin{align}
\label{eq24}
m_{\rm f}\left(r_\times\left(v\right)\right)
= m_{\rm i}\left(r_\times\left(v\right)\right)
+\mathcal{O}\left[\left(v-v_*\right)^{-3/2} \right],
\end{align}
which gives $m_{\rm f}\left(r_\times\left(v\right)\right)-m_{\rm i} \left(r_\times\left(v\right)\right) \to 0$. At the inner horizon, $f_{\rm i}\left(r_-\right)=0$, so the  definition of Misner-Sharp mass yields $m_{\rm i}\left(r_-\right)=r_-/2$. The resulting late-crossing limit is
\begin{align}
\label{eq25}
\lim_{r_\times\to r_-} m_{\rm f}(r_\times)
= m_{\rm i}\left(r_-\right) = \frac{r_-}{2}.
\end{align}
Within the DTR description, the future Misner-Sharp mass remains finite and approaches a scale fixed by the inner-horizon radius. The same $r_-$ also controls the global curvature bound, connecting the shell-crossing dynamics with the geometric structure of the spacetime.

\subsection{Test based on the Ori model}
\label{sec3-3}

We now turn to the Ori model, in which the outgoing perturbation is described by a null shell, while the ingoing perturbation is modeled as a continuous stream of infalling radiation~\cite{Ori1991}. This setup provides a complementary test of whether the present spacetime develops the standard exponential mass inflation near the inner horizon. The
outgoing null shell divides the spacetime into two regions, denoted by $R_1$ and $R_2$. The ingoing radiation profile is prescribed in $R_1$, while the mass function in $R_2$ is determined by the shell matching condition. The metrics in the two regions are written in advanced Eddington-Finkelstein coordinates as
\begin{align}
\label{eq26}
{\rm{d}}s_{1/2}^2 =  - {f_{1/2}}\left( {v,r} \right){\rm{d}}{v^2} + 2{\rm{d}}v {\rm{d}}r + {r^2}{\rm{d}}{\Omega ^2},
\end{align}
with the corresponding masses as ${m}_{1/2}\left ( v,r\right )=r [1-{f}_{1/2}\left ( v,r\right )  ]/2$. Here, $f_1$ and $m_1$ refer to the metric function and Misner-Sharp mass in region $R_1$, respectively, whereas $f_2$ and $m_2$ denote the corresponding quantities in region $R_2$.   We assume that both $f_1$ and $f_2$ have the same functional form in  Eq.~(\ref{eq6}), with the mass parameter promoted to time-dependent functions $M_1\left(v\right)$ and $M_2\left(v\right)$, respectively.

Along the outgoing shell trajectory $r=R\left(v\right)$, the null condition
associated with Eq.~(\ref{eq26}) gives ${{\rm d}R}/{{\rm d} v}=f_1\left(v,R\left(v\right)\right)/2$. This is the same outgoing-null-ray equation considered in Sec.~\ref{sec3-1}. Assuming that $M_1\left(v\right)$ approaches a finite late-time value, the near-horizon scaling derived there applies to the shell trajectory at leading order,
\begin{subequations}
\label{eq30}
\begin{align}
R\left(v\right)-r_- &\simeq \alpha_-^{-1/2}
\left(v-v_*\right)^{-1/2},
\\
\left|f_1\left(v,R(v)\right)
\right|^{-1} &\simeq \alpha_-^{1/2}
\left(v-v_*\right)^{3/2},
\quad
v\to\infty,
\end{align}
\end{subequations}
where $\alpha_-$ is evaluated for the late-time background. The geometric blueshift factor $\left|f_1\right|^{-1}$ grows only as a power law, in contrast to the exponential growth associated with a nondegenerate Cauchy horizon. Eq.~(\ref{eq30}) determines both the late-time position of the outgoing shell and the value of $f_1$ evaluated along its trajectory. These quantities provide the near-horizon input needed to relate the mass evolution on the two sides of the shell.

To determine how the near-horizon behavior in Eq.~(\ref{eq30}) affects the internal Misner-Sharp mass, we impose the matching condition across the outgoing shell~\cite{CarballoRubio2021},
\begin{align}
\label{eq31}
\frac{1}{f_1} \left. \frac{\partial m_1}{\partial v} \right|_{r=R\left(v\right)}
= \frac{1}{f_2} \left.
\frac{\partial m_2}{\partial v}
\right|_{r=R\left(v\right) }.
\end{align}
The metric function $f_1\left(v,R\left(v\right)\right)$ appearing in this relation is evaluated along the same shell trajectory described by Eq.~(\ref{eq30}). The matching condition therefore relates the prescribed mass evolution in $R_1$ to the internal mass evolution in $R_2$. Once $M_1\left(v\right)$ is specified, the outgoing-shell equation, Eq.~(\ref{eq31}), and the mass definitions above determine the coupled evolution of $R\left(v\right)$ and $m_2\left(v,R\left(v\right)\right)$.

For the numerical illustration, we describe the ingoing radiation by the standard Price-law tail~\cite{Price:1971fb,Price:1972pw},  ${M_1}\left( v \right) = {M_0} - \beta {\left( {{v \mathord{\left/ {\vphantom {v {{v_0}}}} \right.  \kern-\nulldelimiterspace} {{v_0}}}} \right)^{ - \gamma }}$, which gives ${\rm d}M_1/{\rm d}v\sim v^{-\gamma-1}$ at late times.  We choose $\gamma=12$ and numerically solve the outgoing-shell equation together with the matching condition. Eq.~(\ref{eq30}) characterizes the local  geometric amplification near the inner horizon, while the behavior of $m_2$ is determined from the coupled numerical evolution.

\begin{figure}[htbp]
\centering
\includegraphics[width=.45\textwidth,origin=c,angle=0]{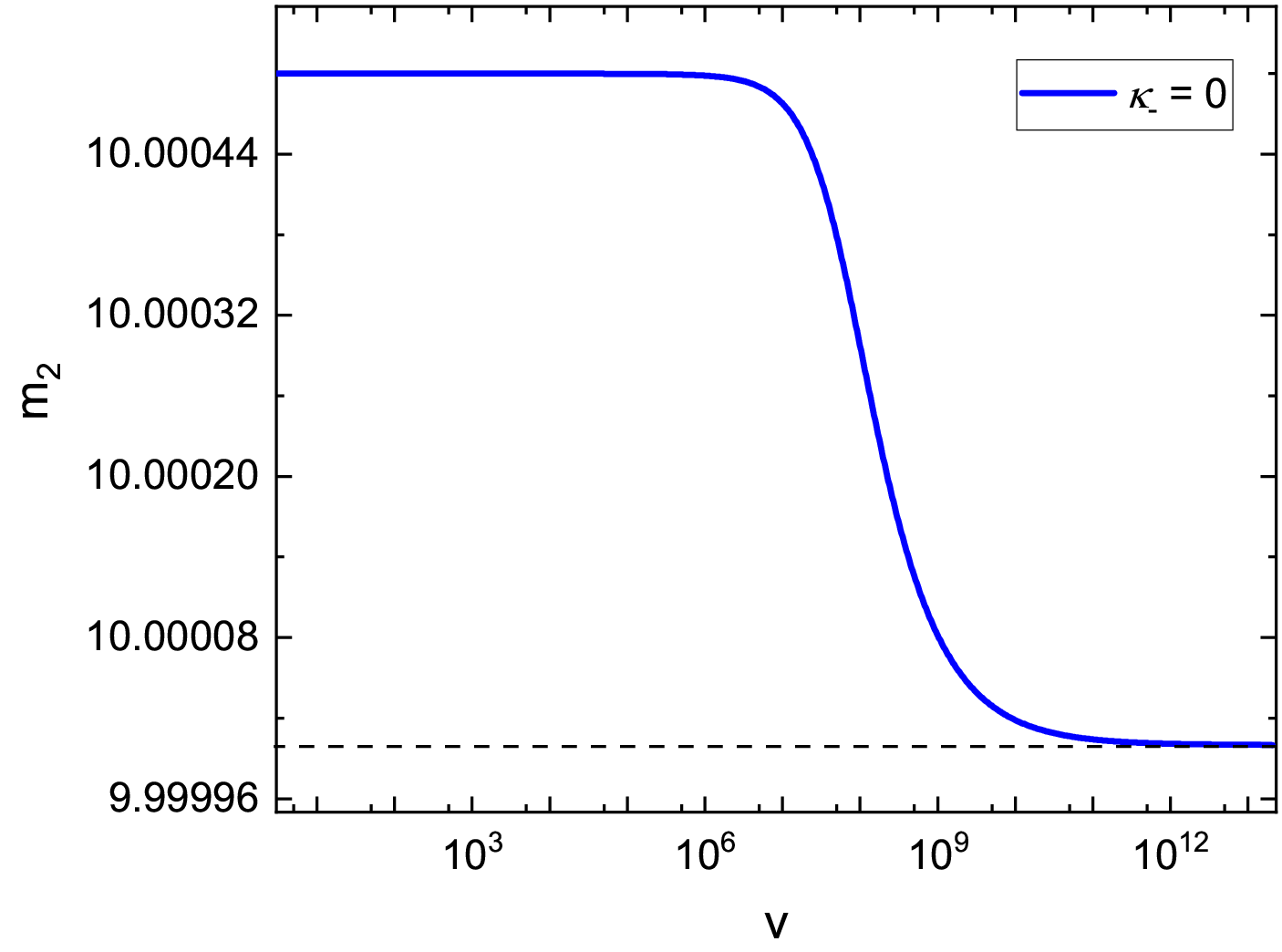}
\caption{\label{fig4} Numerical evolution of the Misner-Sharp mass $m_2$ along the outgoing shell in the Ori model for the $\kappa_-=0$ inner horizon. We choose $M_0=10^6$, $r_-=20$, $r_+=2M_0$, and $\gamma=12$.}
\end{figure}
Figure~\ref{fig4} shows that, for the inner horizon with $\kappa_-=0$, the Misner-Sharp mass ${m_2}\left(v,R\left(v\right)\right)$ remains bounded and approaches a constant value at late times. This plateau has a simple geometric origin. Along the outgoing shell, $R\left(v\right)\to r_-$ and ${f_2}\left(v,R\left(v\right)\right)\to0$, so that
\begin{align}
\label{eq32}
{m_2}\left( {v,R\left( v \right)} \right) = \frac{{R\left( v \right)}}{2}\left[ {1 - {f_2}\left( {v,R\left( v \right)} \right)} \right] \to \frac{{{r_ - }}}{2}.
\end{align}
For $r_-=20$, Eq.~(\ref{eq32}) gives $m_2\to 10$, in agreement with the numerical result. The Ori evolution exhibits no unbounded growth of the internal Misner-Sharp mass, and its late-time value is determined by the degenerate inner-horizon radius. This bounded behavior is consistent with the geometric blueshift of the power-law in Eq.~(\ref{eq30}), in contrast to the exponential amplification associated with a nondegenerate Cauchy horizon. Within the Ori description considered here, the standard exponential mass inflation behavior is not observed. The DTR analysis in Sec.~\ref{sec3-2} gives the same finite late-time mass scale, providing a complementary effective test of the cubic near-horizon behavior.

\subsection{Comparison with representative regular black hole models}
\label{sec3-4}
The results above place the present  construction within the broader class of regular black hole models. Table~\ref{tab1}
compares the core geometry, the inner-horizon structure, the control of the curvature scale, and the corresponding near-horizon behavior. The comparison is intended to summarize the properties relevant to the present analysis rather than to provide an exhaustive classification.

\begin{table*}[htbp]
\centering
\caption{Schematic comparison of  representative regular black hole models, focusing on the properties relevant to the present analysis.}
\label{tab1}
\resizebox{\textwidth}{!}{
\begin{tabular}{lcccc}
\hline
Model
& Core
& Inner horizon
& Curvature control
& Near-horizon behavior
\\
\hline

Bardeen/Hayward-type RBHs~\cite{Bardeen1968,Hayward2006}
& de Sitter-like
& typically nondegenerate, $\kappa_-\neq0$
& set by regularization parameters
& generic exponential amplification
\\

Representative inner-extremal RBHs~\cite{CarballoRubio2022,Franzin2022}
& model-dependent
& degenerate, $\kappa_-=0$
& model-dependent
& exponential factor softened
\\

Present model
& Minkowski
& effective cubic zero, $\kappa_-=0$
& mainly controlled by $r_-$
& $|f|^{-1}\sim v^{3/2}$, bounded $m$ in the DTR and Ori tests
\\
\hline
\end{tabular}
}
\end{table*}

The present model belongs to the inner-extremal class and shares the softening of the standard exponential amplification associated with $\kappa_-=0$. Its model-specific feature is the simultaneous realization of a Minkowski core, an effective cubic inner-horizon zero, and a global curvature scale controlled mainly by $r_-$. The same inner-horizon radius also determines the finite late-time mass scale obtained in the DTR and Ori  descriptions. The comparison shows that the present geometry connects the
global curvature control with a cubic near-horizon mass response, while the same scale $r_-$ fixes the finite late-time mass  obtained in both effective shell descriptions.

\section{Conclusions}
\label{sec4}
In this work, we have constructed a static spherically symmetric regular black hole with an asymptotically Schwarzschild exterior, a regular Minkowski core, a non-extremal outer horizon, and a degenerate inner horizon with $\kappa_-=0$. A key ingredient of the construction is the auxiliary function $g\left(r\right)$, which is chosen to have a double zero at $r=r_-$. As a result, the metric function develops an effective cubic zero at the inner horizon. This ensures the vanishing inner horizon surface gravity and the required sign change of the metric function, while the spacetime remains regular at the center and approaches a Minkowski core as $r\to0$.

We found that in the large-mass regime with $r_+=2M$, the curvature becomes nearly independent of the ADM mass and is mainly controlled by the radius of the inner horizon $r_-$. For the cases considered, the maximum curvature lies inside the inner horizon $r_{\max}<r_-$. In Planck units, the condition $r_-\gtrsim 6.10174$ is sufficient to ensure $K_{\max}<1$, so that the Kretschmann scalar remains sub-Planckian everywhere. We also examined the effective stress-energy tensor reconstructed from the geometry. The radial NEC is identically saturated, while the most significant violation of the angular NEC occurs within a finite region below the inner horizon and becomes strongly suppressed away from this region.

Finally, we have examined the near-inner-horizon behavior using three complementary probes. The massless scalar test field shows that wave propagation reflects the cubic near-horizon structure and that the geometric blueshift factor grows as $\left(v-v_*\right)^{3/2}$ rather than exponentially. We then applied this local scaling to the DTR and Ori models. In the DTR description, the first nonvanishing mass response at fixed $r_-$ is cubic in $r-r_-$, and the nonlinear shell-crossing contribution decays as $\left(v-v_*\right)^{-3/2}$ at late times for finite shell amplitudes, while the future Misner-Sharp mass approaches $m_{\rm f}\to r_-/2$. In the Ori model, the numerical evolution with a Price-law ingoing flux shows that $m_2\left(v,R\left(v\right)\right)$ remains bounded and approaches the same finite scale $r_-/2$. Within these effective descriptions, the cubic degeneracy of the inner horizon replaces the standard exponential amplification with power-law behavior and prevents the corresponding Misner-Sharp masses from growing without bound.

The present analysis treats $r_-$ as an independent fixed inner-horizon scale. If $r_-$ depends on $M$, the total mass response acquires additional terms that can modify the near-horizon power laws. Such mass-dependent families require a separate analysis.

\begin{acknowledgements}
We are very grateful to Binye Dong, Kai Li, Xiaoning Wu, and Zhangping Yu for helpful discussions.
\end{acknowledgements}

\appendix
\section{Full expression of Ricci scalar, the quadratic Ricci invariant, and the Kretschmann scalar}
\label{appA}
In a spherically symmetric static spacetime, 
the Ricci scalar is 
\begin{align}
\label{A1}
R(r)& =-f^{\prime\prime}(r)-\frac{4f^{\prime}(r)}{r}+\frac{2[1-f(r)]}{r^2}
\nonumber \\
&=\mathcal{R}_1+\mathcal{R}_2+\mathcal{R}_3,
\end{align}
according to the metric function~(\ref{eq6}), one has
\begin{subequations}
\label{A2}
\begin{align}
\mathcal{R}_1 &= -\frac{2\mathcal{A}M}{\left[2\mathcal{A}Mr+g\left(r-r_-\right)\left(r-r_+\right)\right]^3}
\nonumber\\
&\times \left\{ g\left[4\mathcal{A}Mr_-r_+ +r\left(r-r_-\right)^2\left(r-r_+\right)^2g'' \right. \right.
\nonumber\\
&\left. -2\left(r-r_-\right)\left(r^2-r_-r_+\right)\left(r-r_+\right)g'\right]
\nonumber\\
&+2r\Big[\mathcal{A}Mr\left(r-r_-\right)\left(r-r_+\right)g''+2\mathcal{A}M
\nonumber\\
&\times \left. \left(r^2-r_-r_+\right)g'-\left(r-r_-\right)^2\left(r-r_+\right)^2\left(g'\right)^2\right]
\nonumber\\
&\left. -2g^2\left[r^3+r_-r_+\left(r_-+r_+-3r\right)`\right]\right\}.
\\
\mathcal{R}_2&=-\frac{8\mathcal{A}M}{r\left[2\mathcal{A}Mr+g\left(r-r_{-}\right)\left(r-r_{+}\right)\right]^2}
\nonumber \\
&\times\left[r\left(r-r_{-}\right)\left(r-r_{+}\right)g^{\prime}+g\left(r^2-r_{-}r_{+}\right)\right].
\\
\mathcal{R}_3&=\frac{4\mathcal{A}M}{r\left[2\mathcal{A}Mr+g\left(r-r_{-}\right)\left(r-r_{+}\right)\right]}.
\end{align}
\end{subequations}
with $\mathcal{A} =  e - 1$, $g'\left( r \right) \!= \!\frac{{2{r_ - }\left( {r - {r_ - }} \right)}}{{{r^3}}}\exp {\left( {\frac{{{r_ - }}}{r} - 1} \right)^2}$, and $g''\left( r \right) \! = \! \frac{{2{r_- }}}{{{r^6}}}\{ {r_- }\left[ {5{r^2}\! + \!2{r_- }\left( {{r_- }\! -\! 2r} \right)} \right] - 2{r^3}\} \exp {\left( {\frac{{{r_- }}}{r} \! - \!1} \right)^2}$. 

The expression of quadratic Ricci invariant is
\begin{align}
\label{A3}
R_{\mu\nu}R^{\mu\nu}&\! = \! 2\left[\frac{f''\left(r\right)}{2}\!+\!\frac{f'\left(r\right)}{r}\right]^2\!+\!\frac{2\left[1\!-\! f\left(r\right)\! - \! rf'\left(r\right)\right]^2}{r^4}
\nonumber\\
&=\mathcal{P}_1+\mathcal{P}_2.
\end{align}
where
\begin{subequations}
\label{A4}
\begin{align}
\mathcal{P}_1&=\frac{2\mathcal{A}^2M^2}
{r^2\left[2\mathcal{A}Mr+g\left(r-r_-\right)\left(r-r_+\right)\right]^6}
\nonumber\\
&\times\left\{r^2\left[\left(r-{r_-}\right)\left(r- {r_+}\right)g''+2 \left(2r-{r_-}-{r_+}\right) \right. \right.
\nonumber\\
&\left. \times g' +2g\right] \left[2\mathcal{A}Mr +g\left(r-{r_-}\right)\left(r-{r_+}\right) \right]
\nonumber\\
&\left. \!- \!2\left[r\left(r\!-\!{r_-}\right)\left(r\!-\!{r_+}\right)g' +g\left(r^2\!-\!{r_-}{r_+}\right)\right]^2 \right\}^2\!.
\\
\mathcal{P}_2&= \frac{8\mathcal{A}^2M^2}
{r^2\left[2\mathcal{A}Mr+g\left(r-r_-\right)\left(r-r_+\right)\right]^4}
\nonumber\\
&\times \left[2\mathcal{A}Mr+g\left(r-{r_-}\right)\left(r-{r_+}\right) \right.
\nonumber\\
&\left.\! - \!r\left(r\!-\!{r_-}\right)\left(r-{r_+}\right)g' \!- \!g \left(r^2\!-\!{r_-}{r_+}\right) \right]^2.
\end{align}
\end{subequations}

The Kretschmann scalar can be simply expressed as
\begin{align}
\label{A5}
K\left( r \right)& = {\left[ {f''\left( r \right)} \right]^2} + \frac{{4{{\left[ {f'\left( r \right)} \right]}^2}}}{{{r^2}}} + 4\frac{{{{\left[ {1 - f\left( r \right)} \right]}^2}}}{{{r^4}}}
\nonumber\\
&=\mathcal{K}_1+\mathcal{K}_2+\mathcal{K}_3,
\end{align}
where
\begin{subequations}
 \label{A6}   
\begin{align}
{{\mathcal{K}}_1}& \!= \!4{{\mathcal{A}}^2}{M^2}\left\{ {g\left[ {4{\mathcal{A}}M{r_- }{r_+ }\! + \!r{{\left( {r\! -\! {r_- }} \right)}^2}{{\left( {r\! -\! {r_ + }} \right)}^2} } \right.} \right.
\nonumber\\
&\left. {\times g''- 2\left( {r - {r_ - }} \right)\left( {{r^2} - {r_ - }{r_ + }} \right) \left( {r - {r_ + }} \right)g'} \right]
\nonumber\\
& + 2r\left[ {{\mathcal{A}}Mr\left( {r - {r_ - }} \right)\left( {r - {r_ + }} \right)g'' + 2{\mathcal{A}}M } \right.
\nonumber\\
&\times \left( {{r^2} - {r_ - }{r_ + }} \right)g' -{{\left( {r - {r_ - }} \right)}^2}\left. { {{\left( {r - {r_ + }} \right)}^2}{{g'}^2}} \right]
\nonumber\\
&{\left. -2g^2[r^3+{r_-}{r_+}({r_-}+{r_+}-3r)] \right\}}/[2\mathcal{A} M r
\nonumber\\
&+g(r-{r_-})(r-{r_+})]^6,
 \\
\mathcal{K}_2 & \!= \!\frac{{16{\mathcal{A}^2}{M^2}{{\left[ {r\left( {r - {r_ - }} \right)\left( {r - {r_ + }} \right)g' + g\left( {{r^2} - {r_ - }{r_ + }} \right)} \right]}^2}}}{{{r^2}{{\left[ {2\mathcal{A}Mr + g\left( {r - {r_- }} \right)\left( {r - {r_ + }} \right)} \right]} ^4}}}\! ,
 \\
\mathcal{K}_3 & =\frac{{16{\mathcal{A}^2}{M^2}}}{{{r^2}{{\left[ {2\mathcal{A}Mr + g\left( {r - {r_ - }} \right)\left( {r - {r_ + }} \right)} \right]}^2}}}.
\end{align}
\end{subequations}
In the limit $r\to0$,  $R$, $R_{\mu\nu}R^{\mu\nu}$ and $K$ vanish, indicating the spacetime approaches a Minkowski core near the center.

\section{Large-mass expansion of the peak curvature}
\label{appB}
In this appendix, we derive the large-mass expansion of the global maximum of the Kretschmann scalar. The higher-order corrections can be treated perturbatively, so that the position of the maximum is shifted from $u_0$ to
\begin{align}
\label{B1}
{u_*}(\varepsilon ) = {u_0} + \varepsilon {u_1} + {\varepsilon ^2}{u_2} +  \cdots,
\end{align}
where the coefficients $u_i$ are determined by the extremum condition  ${{{\rm{d}}K\left( {u,\varepsilon } \right)} \mathord{\left/ {\vphantom {{{\rm{d}}K\left( {u,\varepsilon } \right)} {{\rm{d}}u}}} \right. \kern-\nulldelimiterspace} {{\rm{d}}u}} = 0$. Substituting Eq.~(\ref{B1}) into the extremum condition and expanding order by order in $\varepsilon$,  one finds that the $\mathcal{O}(\varepsilon^0)$ term reproduces ${{A'}_0}\left( {{u_0}} \right) = 0$, while the $\mathcal{O}(\varepsilon^1)$ term gives
\begin{align}
\label{B2}
 {{A''}_0}({u_0}){u_1} + {{A'}_1}({u_0}) = 0 \Rightarrow {u_1} =  - \frac{{{{A'}_1}({u_0})}}{{{{A''}_0}({u_0})}},
\end{align}
and the  corresponding global maximum is then given by evaluating $K$ at the shifted position as
\begin{align}
\label{B3}
{K_{\max }}\left( \varepsilon  \right) & = {K}\left( {{u_*}(\varepsilon ),\varepsilon } \right) 
\nonumber\\
& \!=\! \frac{1}{{r_ - ^4}}\!\left[ {{A_0}\!\left( {{u_*}(\varepsilon )} \right) \!+ \!\varepsilon {A_1}\!\left( {{u_*}(\varepsilon )} \right) \! + \!{\mathcal{O}}\left( {{\varepsilon ^2}} \right)} \right]\!.
\end{align}
Since $\varepsilon$ is small, one may expand
\begin{align}
\label{B4}
A_{0}\left(u_{*}(\varepsilon)\right)&=A_{0}\left(u_{0}+\varepsilon u_{1}+\cdots\right)
\nonumber\\
&=A_{0}\left(u_{0}\right)+\varepsilon u_{1} A_{0}^{\prime}\left(u_{0}\right)+\mathcal{O}\left(\varepsilon^{2}\right),
\\
\varepsilon A_{1}\left(u_{*}(\varepsilon)\right)&=\varepsilon A_{1}\left(u_{0}+\varepsilon u_{1}+\cdots\right)
\nonumber\\
&=\varepsilon\left[A_{1}\left(u_{0}\right)+\mathcal{O}(\varepsilon)\right]
\nonumber\\
&=\varepsilon  A_{1}\left(u_{0}\right)+\mathcal{O}\left(\varepsilon^{2}\right).
\end{align}
Substituting the above expressions into~(\ref{B3}), one has
\begin{align}
\label{B5}
{K_{\max }}(\varepsilon ) \!&= \! \frac{1}{{r_ - ^4}} \! \left[ {{A_0}\! \left( {{u_0}} \right) + \varepsilon {u_1}{{A'}_0}\! \left( {{u_0}} \right) + \varepsilon {A_1}\left( {{u_0}} \right) + {\mathcal{O}}\! \left( {{\varepsilon ^2}} \right)} \right]
\nonumber\\
&\!= \!\frac{1}{{r_ - ^4}}\left[ {{A_0}\left( {{u_0}} \right) + \varepsilon {A_1}\left( {{u_0}} \right) +{\mathcal{O}}\left( {\varepsilon ^2} \right)} \right],
\end{align}
where in the second line we have used the extremum condition $A_0'\left(u_0\right)=0$.

\section{Lower bound on the radius of inner horizon $r_-$}
\label{appC}
Since $K_{\max}$ denotes the global maximum of the Kretschmann scalar, the condition $K_{\max}\leq 1$ guarantees that $K\left(r\right)\leq1$ for all $r\in[0,\infty)$. By using ${K_{\max }} = {{{A_{\max }}} \mathord{\left/ {\vphantom {{{A_{\max }}} {r_ - ^4}}} \right. \kern-\nulldelimiterspace} {r_ - ^4}}$, where $A_{\max}= A(u_*,\varepsilon)$ is the value of $A\left(u,\varepsilon\right)$ evaluated at the maximum position $u_*$, the lower bound is
\begin{align}
\label{C1}
{r_{ - \left( {{\rm{lower}}} \right)}} \geq A_{\max }^{1/4}.
\end{align}
Using the result derived in Appendix~\ref{appB}, one finds that $A_{\max}$ is independent of $M$ at leading order and receives only subleading corrections suppressed by inverse powers of $M$. Therefore, the lower bound in Eq.~(\ref{C1}) is expected to exhibit only weak mass dependence in the large-mass regime.

\begin{figure}[htbp]
\centering 
\includegraphics[width=.45\textwidth,origin=c,angle=0]{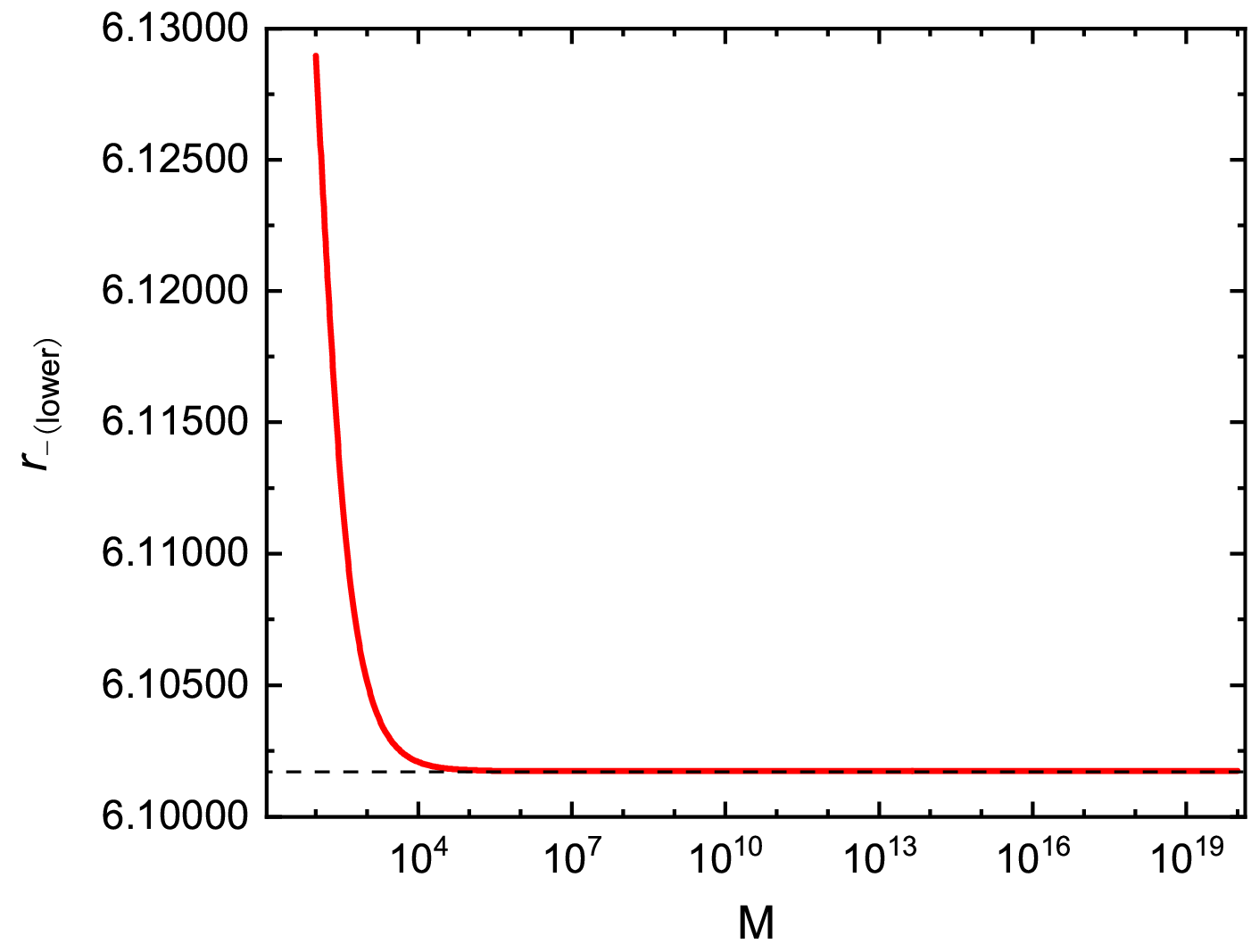}
\caption{\label{fig3} The lower bound of ${r_{ - \left( {{\rm{lower}}} \right)}}$ as a function of $M$. In the large-mass regime, ${r_{ - \left( {{\rm{lower}}} \right)}}  \approx 6.10174$.}
\end{figure}
To verify this behavior, we numerically compute $r_{-(\rm{lower})}$ as a function of $M$, as shown in Fig.~\ref{fig3}. For $M<10^5$, the lower bound exhibits a mild mass dependence, indicating that the subleading $M$-dependent corrections are still non-negligible in this regime. As $M$ increases, $r_{-(\rm{lower})}$ quickly approaches a constant value. In particular, once $M>10^6$, the lower bound becomes essentially independent of the mass and converges to $r_{-(\rm{lower})}\simeq 6.10174$. Therefore, in the large-mass regime considered here, choosing $r_->r_{-(\rm{lower})}$ is sufficient to ensure that the spacetime remains everywhere below the Planck scale. The numerical value $r_{-(\rm{lower})}\simeq6.10174$ should be understood within the parametrization used in this work, namely $r_+=2M$, with $r_-$ treated as an independent inner horizon scale, and in Planck units. It is therefore a sufficient numerical bound for the subclass considered here, rather than a universal lower bound for arbitrary RBH families.

\bibliographystyle{spphys}
\bibliography{template}


\end{document}